\documentclass[structabstract]{aa}
\usepackage{graphicx}
\usepackage{natbib}
\usepackage{txfonts}

\defcitealias{BP1982}{BP}

\begin{document}

   \titlerunning{Modeling the $\mathrm{H}\alpha$ line emission around CTTS}

   \title{Modeling the $\mathrm{H}\alpha$ line emission around classical T Tauri stars using magnetospheric accretion
   and disk wind models}

   \author{G. H. R. A. Lima\inst{1}\thanks{E-mail: styx@fisica.ufmg.br}\and
	  S. H. P. Alencar\inst{1} \and
	  N. Calvet\inst{2} \and
	  L. Hartmann\inst{2} \and
	  J. Muzerolle\inst{3}
	  }

   \institute{Departamento de F{\'{\i}}sica, ICEx-UFMG, CP 702, Belo Horizonte, MG 30123-970, Brazil;\\
	      \and
	      Department of Astronomy, University of Michigan, 830 Dennison Building, 500 Church Street, 
	      Ann Arbor, MI 40109, USA;\\ 
	      \and
	      Space Telescope Science Institute, 3700 San Martin Dr., Baltimore, MD 21218, USA}

   \date{}

\abstract{Spectral observations of classical T Tauri stars show a wide range of line profiles, many of which reveal signs of 
matter inflow and outflow. H$\alpha$ is the most commonly observed line profile owing to its intensity, and it is highly dependent on 
the characteristics of the surrounding environment of these stars.}
{Our aim is to analyze how the H$\alpha$ line profile is affected by the various parameters of our model, which contains both the 
magnetospheric and disk wind contributions to the H$\alpha$ flux.}
{We used a dipolar axisymmetric stellar magnetic field to model the stellar magnetosphere, and a modified Blandford \& Payne model was used 
in our disk wind region. A three-level atom with continuum was used to calculate the required hydrogen level populations. 
We used the Sobolev approximation and a ray-by-ray method to calculate the integrated line profile. Through an extensive study of 
the model parameter space, we investigated the contribution of many of the model parameters on the calculated line profiles.}
{Our results show that the H$\alpha$ line is strongly dependent on the densities and temperatures inside the magnetosphere and the
disk wind region. The bulk of the flux comes most of the time from the magnetospheric component for standard classical T Tauri star 
parameters, but the disk wind contribution becomes more important as the mass accretion rate, the temperatures, and the densities inside
the disk wind increase. We also found that most of the disk wind contribution to the H$\alpha$ line is emitted at the
innermost region of the disk wind.}
{Models that take into consideration both inflow and outflow of matter are a necessity to fully understand
and describe classical T Tauri stars.}

\keywords{Accretion, accretion disks -- Line: profiles -- {\it Magnetohydrodynamics} (MHD) -- Radiative transfer -- 
Stars: formation -- Stars: pre-main sequence}

\maketitle

\section{Introduction}

Classical T Tauri stars (CTTS) are young low-mass stars (\mbox{$\leq 2\mathrm{M}_{\odot}$}), with typical spectral types between
F and M, which are still accreting material from a circumstellar disk. These stars show emission lines ranging from the X-ray
to the IR part of the spectrum, and the most commonly observed one is the H$\alpha$ line. 
Today, the most accepted scenario used to explain the accretion phenomenon in these young stellar objects is the magnetospheric accretion
(MA) mechanism. In this scenario, a strong stellar magnetic field truncates the circumstellar disk near the co-rotation radius,
and the matter in the disk inside this region free-falls onto the stellar photosphere following magnetic
field lines. When the plasma reaches the end of the accretion column, the impact with the stellar surface creates a 
hot spot, where its kinetic energy is thermalized \citep[e.g.,][]{Camenzind1990,Koenigl1991}. The gas is heated to temperatures
of $\simeq 10^{6}\,$K in the shock, which then emits strongly in the X-rays. Most of the X-rays are then reabsorbed by the
accretion column, and re-emitted as the blue and ultraviolet continuum excess \citep[e.g.,][]{CalvetGullbring1998}, 
which can be observed in CTTS.

The magnetospheric accretion paradigm is currently strongly supported by observations. Magnetic field measurements in
CTTS yielded strengths on their surfaces on the order of $10^{3} \mathrm{G}$ persistently over time scales of years
\citep{Johns-Krull_etal1999, Symington_etal2005, Johns-Krull2007}. These fields are strong
enough to disrupt the circumstellar disk at some stellar radii outside the stellar surface. Classical T Tauri stars also exhibit
inverse P Cygni (IPC) profiles, mostly in the Balmer lines that arise from the higher excitation levels, 
and some metal lines which are directly linked to the MA phenomenon \citep{Edwards_etal1994}. These lines are broadened 
to velocities on the order of hundreds of $\mathrm{km \, s^{-1}}$, which is evidence of \mbox{free-falling} gas from a distance of some
stellar radii above the star. There is also an indication of line profile modulation by the stellar rotation period
\citep[e.g.,][]{Bouvier_etal2007}.  

However, the MA scenario is only partially successful in explaining the features observed in line profiles of CTTS. Observational studies
show evidence of outflows in these objects, which are inferred by the blue-shifted absorption features that can be seen
in some of the Balmer and Na D lines \citep[e.g.,][]{AlencarBasri2000}. Bipolar outflows leaving these objects have been observed
down to scales of $\simeq 1\,\mathrm{AU}$ \citep[e.g.,][]{TakamiBailey2003, AppenzellerBertout2005}. On larger scales,
{\it HST} observations of HH30 \citep[e.g.,][]{Burrows_etal1996} could trace a bipolar jet to within $\lesssim 30\, \mathrm{AU}$
of the star. This suggests a greater complexity of the circumstellar environment, in which the MA model is only a piece of the
puzzle. 

The observed jets are believed to be highly collimated magneto-hydrodynamic disk winds which efficiently extract 
angular momentum and gravitational energy from the accretion disk. \citet[][hereafter BP]{BP1982} were the first to propose the use of
a disk wind to explain the origin of jets from accreting disks around a black hole, and soon after \citet{Pudritz_Norman1983, 
Pudritz_Norman1986} proposed a similar mechanism for the origin of the protostellar jets. \citet{Bacciotti_etal2003}, using high 
resolution spectro-imaging and adaptative optics methods on protostellar jets, observed what seems to be rotational motion inside 
these jets, and also 
an onion-like velocity structure where the highest speeds are closer to the outflow axis. Their work strongly corroborates the idea of 
the origin of jets as centrifugally driven MHD winds from extended regions of their accretion disks. A correlation between jets,
infrared excess, and the accretion process has been observed \citep{Cabrit_etal1990,HartiganEG1995}, which indicates a dependence
between the outflow and inflow of matter in CTTS. Disk wind theories show the existence of a scale relation between the disk accretion
($\dot{M}_\mathrm{acc}$) and the mass loss rates ($\dot{M}_\mathrm{loss}$) \citep{Pelletier_Pudritz1992}. Observations have confirmed 
this relation \citep{Hartmann1998}, and both have agreed that in CTTS the typical value for this relation is $\dot{M}_\mathrm{loss}/
\dot{M}_\mathrm{acc} \simeq 0.1$. Another model widely used to explain the outflows and the protostellar jets is called the X-wind model 
\citep{Shu_etal1994, Cai_etal2008}, which states that instead of originating in a wider region of the accretion disk, the 
outflow originates in a very constrained region around the so called `X-point', which is at the Keplerian co-rotation radius of the 
stellar magnetosphere. 

Many developments in MHD time-dependent numerical models have been made in the last years. \citet{Zanni_etal_2007} were able to 
reproduce the disk-wind lauching mechanism and jet formation from a magnetized accretion disk with a resistive MHD axisymmetric model 
showing the system evolution over tens of rotation periods. They were able to find a configuration in which a 
slowly evolving outflow leaving the inner part of the accretion ring (from $\sim 0.1$ to $\sim 1\,\mathrm{AU}$) was formed,
when using a high disk magnetic resistivity. \citet{Murphy_etal_2010} showed that it was possible to form a steady disk wind
followed by a self-contained super-fast-magnetosonic jet using a weakly magnetized accretion disk. Their numerical solution 
remained steady for almost a thousand Keplerian orbits. Both models help to strengthen the disk wind scenario. \citet{Romanova_etal_2009},
however, showed that it is also possible to numerically simulate the launching of a thin conical wind, which is similar 
in some respects to an X-wind, considering only an axisymmetric stellar dipolar magnetic field around a slowly rotating star. 
A fast axial jet component also appears in the case of a fast rotating star. Their conical wind also shows some degree of 
collimation, but, as stated by \citet{Romanova_etal_2009}, this collimation may not be enough to explain the observed 
well-collimated jets. 

Radiative transfer models were used to calculate some of the line profiles that were observed in CTTS. 
The first of these models that used the MA paradigm, hereafter the HHC model, was proposed by \citet*{HHC1994}, and 
was based on a simple axisymmetric dipolar geometry for the accretion flow, using a two-level atom approximation under the Sobolev 
``resonant co-moving surfaces'' approximation (SA). The model was later improved by the addition of the full statistical equilibrium 
equations in the code \citep*{MCH1998}, and finally including an exact integration of the line profile \citep*{MCH2001}. These models 
included only the magnetospheric and photospheric components of the radiation field and were partially successful in reproducing the 
observed line strengths and morphologies for some of the Balmer lines and Na D lines. In these models, the stellar photosphere and the 
inner part of the accretion disk partially occult the inflowing gas, and these occultations create an asymmetric blue-ward or red-ward 
emission peak, depending on the angle between the system symmetry axis, and the observer's line of sight. The inflowing gas projected
against hot spots on the stellar surface produces the IPC profiles that are sometimes observed. 

\citet{Alencar_etal2005} demonstrated that the observed $\mathrm{H}\alpha$, $\mathrm{H}\beta$, and Na D lines of RW Aur
are better reproduced if a disk wind component arising from the inner rim of the accretion disk was added to these radiative transfer models,
and it became clear that for more accurate predictions, these models should include that component. \citet[][henceforth KHS]{KurosawaHS_2006},
were the first to use a self-consistent model where all these components were included, and could reproduce the wide variety of the
observed $\mathrm{H}\alpha$ line profiles. Still, their cold MHD disk wind component used straight field lines, and some of which
violate the launching conditions required by BP, which state that for an outflow to appear, its launching angle should be $>30\degr$ away
from the rotational axis of the system. Those infringing field lines in the KHS model were always the innermost ones, which, as 
we will show here, are responsible for the bulk of the $\mathrm{H}\alpha$ line profiles in the disk wind region. 

The BP launching condition assumes a cold MHD flow, which is not always the case. In a more general situation, the thermal pressure
term becomes important near the base of the outflow, and should not be neglected during the initial acceleration process, and
thus it is possible to have a stable steady solution even if the launching angle exceeds $30\degr$. The BP self-similar solution, 
although it is a very simple one, is able to describe mathematically and self-consistently the disk-wind launching mechanism and 
the further collimation of the disk wind to a jet by magnetic ``hoop'' stress. The self-similarity hypothesis becomes increasingly 
artificial farther away from the accretion disk, in the region where the jet is formed, but it remains valid near the accretion disk, 
where our calculations were performed.

In this work, we have included a disk wind component in the HHC model, using a modified version of the BP formulation.
We then studied the parameter space of the improved model, while trying to identify how each parameter affected the calculated 
H$\alpha$ profiles. In Sects 2 and 3 we present the magnetospheric accretion, the disk wind and the radiative transfer models
that were used as well as all the assumptions made to calculate our profiles. The model results are shown in Sect. 4, followed 
by a discussion in Sect. 5 and the conclusion in Sect. 6. A subsequent paper will focus on the modeling of the observed 
H$\alpha$ lines for a set of classical T Tauri stars that exhibit different accretion and environmental characteristics. 

\section{Model structure}

\begin{figure}[htb]
 \begin{center}
 \vspace{0.2cm}
 \includegraphics[width=88mm]{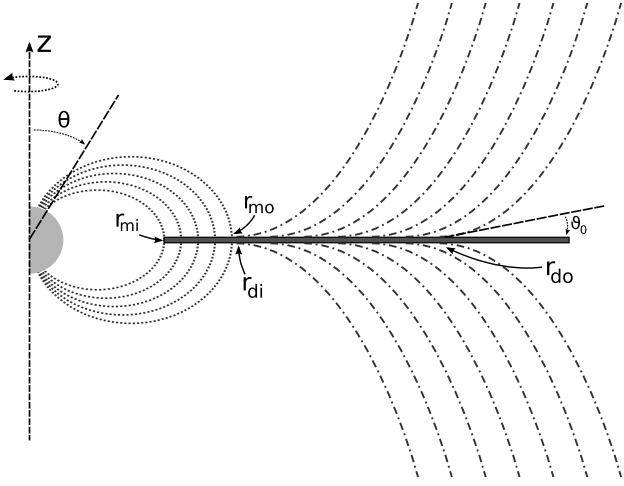}
 \caption{Not-to-scale sketch of the adopted axisymmetric geometry shown in the poloidal plane. The gray half-circle denotes the stellar 
photosphere, the dark rectangle represents the disk, which is assumed to be opaque and to extend to the innermost magnetic field line. The dotted 
lines represent the magnetosphere and its purely dipolar field lines, and the dash-dotted lines are parabolic trajectories
that represent the disk wind streamlines. The inner and outer magnetospheric field lines intersect the disk at $r_\mathrm{mi}$ and 
$r_\mathrm{mo}$, respectively, and the ring between $r_\mathrm{di}$ and $r_\mathrm{do}$ is the region where the disk wind is
ejected. The inclination from the \emph{z-axis} is given by $\theta$, and $\vartheta_0$ is the disk-wind launching angle.}
 \label{CV_Model}
 \end{center}
\end{figure}

Our model structure comprises four components: the star, the magnetosphere, the accretion disk, and the disk wind. The center of the star is
also the center of the model's coordinate system, and the star rotates around the {\it z-axis}, which is the system's axis of symmetry. We
assume symmetry through the {\it x-y plane}. In our coordinate system, $r$ is the radial distance from the center of the star, and
$\theta$ is the angle between $r$ and the {\it z-axis}. The modeled region is divided into a grid system, where $1/4$ of the points are inside the
magnetosphere, and the other $3/4$ are part of the disk wind region. The star and accretion disk are assumed to be infinitely thick in the optical
region of the spectrum, and as such can be said to emit as blackbodies, and act as boundaries in our model. Due to symmetry arguments, it
is only necessary to calculate the velocity, density, and temperature fields in the first quadrant of the model. These values are rotated around
the {\it z-axis}, and are then reflected through the {\it x-y plane}. The emission from the accretion disk in CTTS is neglected, because its 
temperature is usually much lower than the photospheric temperature of the star. A schematic diagram of the adopted geometry can be seen in
Fig.~\ref{CV_Model}.

The stellar surface is divided into two regions: the photosphere and the accretion shock (the region where the accruing gas hits the stellar
photosphere). Unless otherwise stated, the stellar parameters used in the model are those of a typical CTTS, i.e. radius ($\mathrm{R_{*}}$), 
mass ($\mathrm{M_{*}}$) and effective temperature of the photosphere ($T_\mathrm{ph}$) are $2.0\,\mathrm{R_{\odot}}$, $0.5\,\mathrm{M_{\odot}}$ 
and $4000\,\mathrm{K}$, respectively. When the gas hits the star, its kinetic energy is thermalized inside a radiating layer, and we consider
that this layer emits as a single temperature blackbody. Because we assume axial symmetry, our accretion shock region can be regarded as two spherical 
rings near both stellar poles, and we assume their effective temperature ($T_\mathrm{sh}$) to be $8000\,\mathrm{K}$, as used by
\citet{MCH2001}. Although $T_\mathrm{sh}$ has been kept constant, it actually depends on the mass accretion rate. But it barely
affects the H$\alpha$ line profile and is only important when defining the continuum veiling levels.

\subsection{Magnetospheric model}
The basic assumption of this model is that accretion from the disk onto the CTTS is controlled by a dipole stellar magnetic field. This
magnetic field is assumed to be strong enough to truncate the disk at some height above the stellar surface, and also to remain  
undisturbed by the ionized inflowing gas. The gas pressure inside the accretion funnel is assumed to be sufficiently low so
that the infalling material free-falls onto the star.

For the considered axisymmetric flow, the streamlines can be described by

\begin{equation}
 r = r_\mathrm{m} \sin^{2} \theta,
\end{equation}
where $r_\mathrm{m}$ corresponds to the point where the stream starts at the disk surface, or, in other words, where $\theta = \pi/2$ 
\citep{GhoshPethick1977}. In this problem, we can use the ideal magneto-hydrodynamic (MHD) approximation, which assumes an infinite
electrical conductivity inside the flow. According to this approximation, the flow is so strongly coupled to the magnetic field 
lines that the velocity of the inflowing gas is parallel to the magnetic field lines, and thus the poloidal component of the velocity is

\begin{equation}
 \mathbf{v_\mathrm{p}} = - \varv_\mathrm{p} \left[ \frac{3y^{1/2} (1-y)^{1/2} \hat{\varpi} + (2-3y)\hat{z}}{(4-3y)^{1/2}} \right],
\end{equation}
where $y=r/r_\mathrm{m} = \sin^{2} \theta$, $\hat{\varpi}$ is the cylindrical radial direction, and $\varv_\mathrm{p}$ is the gas 
free-fall speed, which according to \citet{HHC1994} can be written as

\begin{equation}
 \varv_\mathrm{p} = \left[ \frac{2GM_\mathrm{*}}{R_\mathrm{*}} \left(\frac{R_\mathrm{*}}{r} - \frac{R_\mathrm{*}}{r_\mathrm{m}} 
\right) \right]^{1/2}.
\end{equation}

This model constrains the infalling gas between two field lines, which intersect the disk at distances $r_\mathrm{mo}$, for the 
outermost line, and $r_\mathrm{mi}$, 
for the innermost one (Fig.~\ref{CV_Model}). The area filled by the accretion hot spots directly depends on these values, because it is bounded by 
the limiting angles $\theta_\mathrm{i}$ and $\theta_\mathrm{o}$, the angles where the lines crossing the disk at $r_\mathrm{mi}$ and 
$r_\mathrm{mo}$ hit the star, respectively. We have set these values to be $r_\mathrm{mi}$=2.2$\,R_\mathrm{*}$ 
and $r_\mathrm{mo}$=3.0$\,R_\mathrm{*}$, which are the same values used in the \mbox{`small \slash wide'} case of \citet{MCH2001}. 
The corresponding accretion ring covers an area of $8\%$ of the total stellar surface, which is on the same order of magnitude as 
the observational estimates \citep{Gullbring_etal1998, Gullbring_etal2000}.
\citet{HHC1994} showed that the gas density $(\rho)$ can be written as a function of the mass accretion rate $\dot{M}_\mathrm{acc}$ onto 
these hot spots as 

\begin{equation}
 \rho = \frac{\dot{M}_\mathrm{acc}}{4\pi (1/r_\mathrm{mi} - 1/r_\mathrm{mo})} \frac{r^{-5/2}}{(2GM)^{1/2}} \sqrt{\frac{4-3y}{1-y}}. 
\end{equation}

\citet{Martin1996} presented a self-consistently solved solution for the thermal structure inside the dipolar magnetospheric accretion funnel
by solving the heat equations coupled to the rate equations for hydrogen. 
When his temperature structure was used by \citet{MCH1998} to compute CTTS line profiles, their results did not agree with the observations. 
This is not the case with theoretical profiles based on the calculated temperature distribution by
\citet{HHC1994}, which we adopt here. The temperature structure used is computed assuming a volumetric heating rate 
$\propto r^{-3}$ and balancing the energy input with the radiative cooling rates used by \citet{HartmannAE1982}.

We used a non-rotating magnetosphere in this model. The rotational speed inside the magnetosphere is much lower compared
to the free-fall speeds inside the accretion columns. \citet{MCH2001} showed that the differences between profiles calculated with 
a rotating and a non-rotating magnetosphere are minimal if the stellar rotational speed is $\lesssim 20\,\mathrm{km/s}$. 

\subsection{Disk wind model}

We adopted a modified BP model for the disk wind. This model assumes an axisymmetric magnetic field 
treading the disk, uses the ideal MHD equations to solve the outflowing flux leaving the disk, and assumes a cold flow, where the 
thermal pressure terms can be neglected. In the BP model, the wind solution is self-similar and covers the whole disk; in this work,
instead, we use a self-similar solution only inside the region limited by the disk wind inner and outer radii. This solution can be 
specified by the following equations:
\begin{eqnarray}
 \mathbf{\nabla}(\rho\mathbf{v})&=&0, \label{continuidade}\\
 \rho(\mathbf{v\cdot\nabla})\varv_\mathrm{z} &=& -\rho \frac{\partial}{\partial z} \left[ \frac{-GM_\mathrm{*}}{(\varpi^{2}+z^{2})^{1/2}} \right]
-\frac{1}{8\pi} \frac{\partial B^{2}}{\partial z} + \frac{1}{4\pi} (\mathbf{B\cdot\nabla}) B_\mathrm{z}, \label{l_momentum}\\
 \mathbf{v}&=&\frac{k\mathbf{B}}{4\pi\rho} + (\mathbf{\omega \times r}), \label{isorotation}\\
 e&=& \frac{1}{2} \varv^{2} - \frac{GM_\mathrm{*}}{(\varpi^{2}+z^{2})^{1/2}} - \omega \frac{\varpi B_\mathrm{\phi}}{k}, \label{energy}\\
 l&=& \varpi \varv_\mathrm{\phi} - \frac{\varpi B_\mathrm{\phi}}{k} \label{angular_mom},
\end{eqnarray}
where $e$, the specific energy, $l$, the specific angular momentum, $k$, the ratio of mass flux to magnetic flux, and $\omega$, the
angular velocity of the magnetic field line, are constants of motion along a flow line \citep[BP]{Mestel1961}. The variables 
$\mathbf{r}$, $\mathbf{v}$, $\mathbf{B}$ are the radial, the flow velocity, and the magnetic field vectors in
cylindrical coordinates respectively. If we solve the problem for the first flux line, the solution can be scaled for the other lines with 
the scaling factors introduced in BP:
\begin{eqnarray}
 \mathbf{r}&=& \left( \varpi_\mathrm{0}\xi(\chi),\phi,\varpi_\mathrm{0}\chi \right), \label{r_scale}\\
 \mathbf{v}&=& \left( \frac{GM_\mathrm{*}}{\varpi_\mathrm{0}} \right)^{1/2} \left( f\xi'(\chi),g(\chi),f(\chi) \right) \nonumber \\ [-5pt]
  &=& \left( \frac{GM_\mathrm{*}}{\varpi_{1}} \right)^{1/2} \left( \frac{\varpi_\mathrm{0}}{\varpi_\mathrm{1}} \right)^{-1/2} \left( f\xi'(\chi),
 g(\chi),f(\chi) \right), \label{v_scale}\\
 \rho&=&\rho_\mathrm{0} \eta(\chi) \nonumber\\ [-5pt]
  &=& \rho_\mathrm{1} \left( \frac{\varpi_\mathrm{0}}{\varpi_\mathrm{1}} \right)^{-3/2} \eta(\chi) , \label{dens_scale}\\
 \mathbf{B}&=&B_\mathrm{0} \left( b_\mathrm{\varpi}(\chi), b_\mathrm{\phi}(\chi), b_\mathrm{z}(\chi) \right) \nonumber \\ [-5pt]
  &=& B_\mathrm{1} \left( \frac{\varpi_\mathrm{0}}{\varpi_\mathrm{1}} \right)^{-5/4} \left( b_\mathrm{\varpi}(\chi), 
b_\mathrm{\phi}(\chi), b_\mathrm{z}(\chi) \right), \label{B_scale} \\
 \chi&=&z/\varpi_\mathrm{0} \label{z_scale},
\end{eqnarray}
where a zero subscript indicates evaluation at the disk surface ($\chi$=0, $\xi$=1), a subscript 1 means that the quantity 
is evaluated at the fiducial radius $\varpi_{1}$, and a prime denotes differentiation with respect to $\chi$. The functions $\xi(\chi)$,
$f(\chi)$, $g(\chi)$, $\eta(\chi)$, $b_{\varpi}(\chi)$, $b_{\phi}(\chi)$, $b_\mathrm{z}(\chi)$ represent the self-similar solution of the
wind. \citet{BP1982} introduced the dimensionless parameters
\begin{eqnarray}
 \epsilon&=& \frac{e}{\left( GM_{*}/\varpi_{0} \right)} \label{epsilon}, \\
 \lambda&=& \frac{l}{(GM_{*}\varpi_{0})^{1/2}}, \label{lambda}\\
 \kappa&=&  k \left(1+\xi'^{2}_{0}\right)^{1/2} \frac{\left(GM_{*}/\varpi_{0}\right)^{1/2}}{B_{0}} \label{kappa} ,
\end{eqnarray}
with $\epsilon$, $\lambda$, $\kappa$ and $\xi'_{0}$ the constant parameters that define the solution. With the help of 
Eqs.~(\ref{r_scale})$-$(\ref{kappa}), and knowing that the Keplerian velocity is $\omega = (GM/\varpi_{0}^{3})^{1/2}$, 
one can find after some algebraic manipulation \citep[see BP,][]{Safier1993} the following quartic equation for $f(\chi)$,
\begin{equation}
 T - f^{2}U = \left[\frac{(\lambda - \xi^{2})m}{\xi(1-m)}\right]^{2}, \label{quartic}
\end{equation}
where
\begin{eqnarray}
 m&\equiv&\frac{4\pi\rho(\varv^{2}_{\varpi} + \varv^{2}_\mathrm{z})}{B^{2}_{\varpi} + B^{2}_\mathrm{z}} \label{alfvenmach}, \\
 &=& \kappa \xi f J
\end{eqnarray}
is the square of the (poloidal) Alfv\'en Mach number, and
\begin{eqnarray}
 T&=&\xi^{2} + 2S -3, \\
 S&=&(\xi^{2} + \chi^{2})^{-1/2}, \\
 U&=&(1+\xi'^{2}), \\
 J&=&\xi -\chi \xi'.
\end{eqnarray}

The flow trajectory solution is a self-similar solution that always crosses the disk at $\xi=1$. If this self-similar
solution, which is given by $\chi(\xi)$, is expanded into a polynomial function around the disk-crossing point using Taylor's 
theorem, and we consider only the first three terms of the expansion, we will have a parabolic trajectory describing the outflow.
This parabolic trajectory is given by
\begin{equation}
 \chi(\xi) = a\xi^{2} +b\xi + c, ~~~~\mathrm{with}~~~~c=-(a+b),
 \label{flow_eq}
\end{equation}
This solution respects the launching condition, which states that the flow-launching angle $\vartheta_{0}$ (see Fig.~\ref{CV_Model}) 
should be less than $\vartheta_\mathrm{c} = 60\degr$ for the material to accelerate magneto-centrifugally and be ejected as a wind
from the disk. Equation~(\ref{quartic}) can be rewritten as a quartic equation for $m$,
\begin{eqnarray}
 \xi^{2}Vm^{4} - 2\xi^{2}Vm^{3} - \left[ \xi^{2}(T-V) - (\lambda-\xi^{2})^{2} \right] m^{2}+&&  \nonumber \\
 + 2\xi^{2}Tm + \xi^{2}T& =&0,
 \label{quartic_m}
\end{eqnarray}
\begin{equation}
 \mathrm{where} ~~~~~ V = \frac{U}{(\kappa \xi J)^{2}}.
\end{equation}
The solution of Eq.~(\ref{quartic_m}) gives four complex values of $m$ for each pair of points $(\chi,\xi)$ inside the trajectory. The
flow must reach super-Alfv\'enic speeds, for a strong self-generated toroidal field to appear, collimating the flux, and thus producing 
a jet \citep{BP1982}. Of the four solutions of this quartic equation, there are only two real solutions that converge on one 
another at the Alfv\'en point, a sub-Alfv\'enic and a super-Alfv\'enic one. The sub-Alfv\'enic solution reaches the Alfv\'en 
speed and then the solution starts decreasing again, thus remaining a sub-Alfv\'enic solution before and after the Alfv\'en 
point. The same happens with the super-Alfv\'enic solution, which remains super-Alfv\'enic before and after the Alfv\'en point. 
The chosen solution must be a continuous real solution, and must also increase monotonically with $\chi$, because 
the wind is constantly accelerated by the magnetic field. This solution is obtained by using the convergent 
sub-Alfv\'enic solution before the Alfv\'en radius, and then the super-Alfv\'enic one.

The launching condition imposes that $\xi'_{0} \gtrsim \cot \vartheta_\mathrm{c}$, and it helps constrain the set of values of $a$, $b$, and $c$ 
that can be used in equation~(\ref{flow_eq}). Then we can write
\begin{equation}
 \xi'_{0} = \frac{1}{2a+b} \gtrsim \cot \vartheta_\mathrm{c}.
\end{equation}
If the values of parameters $\kappa$ and $\lambda$ are also given, the flow can be fully described by the following equations:
\begin{eqnarray}
 f&=&\frac{m}{\kappa \xi J}, \label{Eq_f} \\
 f_\mathrm{p}&=& f(1+\xi'^{2})^{1/2}, \label{Eq_fp} \\
 g&=&\frac{\xi^{2} -m\lambda}{\xi(1-m)}, \label{Eq_g} \\
 \eta&=& \frac{(\xi f J)_{0}}{\xi f J} = \frac{m_0}{m} \label{Eq_eta},
\end{eqnarray}
where $\eta$ and $f_\mathrm{p}$ are the self-similar dimensionless density and poloidal velocity, respectively. Knowing the gas density $\rho$,
the poloidal velocity $\mathbf{v_\mathrm{p}}$, and the scaling laws, it can be shown that the mass loss rate $\dot{M}_\mathrm{loss}$ 
is given by
\begin{equation}
 \dot{M}_\mathrm{loss} = 2\pi \rho_{0} f_\mathrm{p0} \left( GM_{*} \varpi_{0}^{3} \right)^{1/2} \ln \frac{r_\mathrm{do}}{r_\mathrm{di}},
 \label{mloss}
\end{equation}
where $r_\mathrm{di}$ and $r_\mathrm{do}$ are the inner and outer radii in the accreting disk that delimits the wind ejecting region (see 
Fig.~\ref{CV_Model}). With this equation it is possible to calculate the value of $\rho_{0}$, which is necessary to produce a 
$\dot{M}_\mathrm{loss}$, when the wind leaves the disk between $r_\mathrm{di}$ and $r_\mathrm{do}$.  

The temperature structure we used inside the disk wind is similar to the one used in the magnetospheric accretion funnel. There
are some studies about the temperature structure inside this region \citep{Safier1993, Panoglou_etal2010}, and we are going to
discuss them in Sect. 6.

The parameters used to calculate the self-similar wind solution were initially: $a$=0.43, $b$=$-0.20$, $\lambda$=$30.0$, $\kappa$=$0.03$.
These values of $a$ and $b$ were chosen to produce an intermediate disk-wind lauching angle $\vartheta_{0}$$\simeq$$33\degr$, while for $\lambda$ 
and $\theta$ these values are the same as the ones used by BP in their standard solution. We later changed some of these values to 
see how they influence our calculated H$\alpha$ line profiles. The disk wind inner radius $r_\mathrm{di}$=$3.01\, R_{*}$ is just slightly 
larger than $r_\mathrm{mo}$, and we changed the value of the disk wind outer radius $r_\mathrm{do}$ from $5\, R_{*}$ to $30\, R_{*}$ to 
see how this affected the calculated profiles. To calculate $\rho_{0}$ we used a base value of $\dot{M}_\mathrm{loss}$=$10^{-9} \,
\mathrm{M_{\odot}\,yr^{-1}}$=$0.1\, \dot{M}_\mathrm{acc}$.

\section{Radiative transfer model}

In order to calculate the H$\alpha$ line, it is necessary to know the population of at least levels 1 to 3 of the hydrogen atom. 
The model used for the hydrogen has three levels plus continuum. The high occupation number of the ground level, owing to the 
low temperatures in the studied region, leads to an optically thick intervening medium to the Lyman lines. Using this, 
it can be assumed that the fundamental level is in LTE. The other two levels plus ionization fraction were calculated with 
a two-level atom approximation \citep[see][chapter 11]{Mihalas1978}. The fundamental level population is used to better
constrain the total number of atoms that are in levels 2 and 3, and is necessary to calculate the opacity due to H$^{-}$ ion.

The line source function is calculated with the Sobolev approximation method \citep{Rybicki_Hummer1978A, HHC1994}. This method can
be applied when the Doppler broadening is much larger than the thermal broadening inside the flow. Thus, an atom emitting a specific
line transition only interacts with other atoms having a limited range of relative velocities, which occupy a small volume of the
atmosphere. In the limit of high velocity gradients inside the flow, these regions with atoms that can interact with a given point
additionally become very thin, and it can be assumed that the physical properties across this thin and small region remain constant. These small
and thin regions are called ``resonant surfaces''. Each point inside the atmosphere then only interacts with a few resonant surfaces,
which simplifies the radiative transfer calculations a lot. The radiation field at each point can be described by the mean intensity
\begin{equation}
 \bar{J}_{\nu} = [1 -\beta(\mathbf{r})]S(\mathbf{r}) + \beta_\mathrm{c}(\mathbf{r})I_\mathrm{c} +F(\mathbf{r}),
 \label{avr_intens}
\end{equation}
where $\beta$ and $\beta_\mathrm{c}$ are the scape probabilities of local and continuum (stellar photosphere plus accretion shock,
described by $I_\mathrm{c}$) radiation, $S$ is the local source function, and $F$ is the non-local term, which takes into account only the
contributions from the resonant surfaces. A more detailed description of each of these terms has been given in \citet{HHC1994}.

For a particular line transition, the source function is given by
\begin{equation}
 S_{ul} = \frac{2h\nu^{3}_{ul}}{c^{2}} \left[ \left( \frac{N_{l} g_{u}}{N_{u} g_{l}} \right) -1 \right]^{-1},
 \label{line_source}
\end{equation}
where $N_{l},N_{u},g_{l}$ and $g_{u}$ are the population and statistical weights of upper and lower levels of the
transition respectively.

The flux is calculated as described by \citet{MCH2001}, where a ray-by-ray method was used. The disk coordinate system is rotated
to a new coordinate system, given by the cartesian coordinates $(P,Q,Z)$, in which the $Z$-axis coincides 
with the line of sight, and each ray is parallel to this axis. The specific intensity of each ray is given by
\begin{equation}
 I_{\nu} = I_{0} \mathrm{e}^{-\tau_\mathrm{tot}} + \int^{\tau(Z_{0})}_{\tau(-\infty)} {S_{\nu}(\tau')\mathrm{e}^{-\tau'} \mathrm{d}\tau'}, 
 \label{spec_intens}
\end{equation}
where $I_{0}$ is the incident intensity from the stellar photosphere or the accretion shock, when the rays hit the star, and is zero 
otherwise. $S_{\nu}(Z)$ is the source function at a point outside the stellar photosphere at frequency $\nu$, and $\tau_\mathrm{tot}$
is the total optical depth from the initial point, $Z_{0}$, to the observer ($-\infty$), and $\tau(-\infty) =0$. $Z_{0}$ can be either 
the stellar surface, the opaque disk, or $\infty$. The optical depth from a point at $Z$ to the observer is calculated
as
\begin{equation}
 \tau(Z) = - \int^{Z}_{-\infty} {\left[ \chi_\mathrm{c}(Z') + \chi_\mathrm{l}(Z') \right] \mathrm{d}Z'},
 \label{optical_depth}
\end{equation}
in which $\chi_\mathrm{c}$ and $\chi_\mathrm{l}$ are the continuum and line opacities. The continuum opacity and emissivity contain the contributions
from hydrogen bound-free and free-free emission, \element[-][]{H}, and electron scattering \citep[see][]{Mihalas1978}.

This model uses the Voigt profile as the profile of the emission or absorption lines that were modeled. Assuming a damping constant 
$\Gamma$, which controls the
broadening mechanisms, \mbox{$a=\Gamma /4 \pi \Delta \nu_\mathrm{D}$}, \mbox{$\varv=(\nu - \nu_{0})/\Delta \nu_\mathrm{D}$}, and 
\mbox{$Y=\Delta\nu / \Delta\nu_\mathrm{D}$}, where $\Delta \nu_\mathrm{D}$ is the Doppler broadening due to thermal effects, and 
$\nu_{0}$ the line center frequency, the Voigt function can be written as  
\begin{equation}
 H(a,\varv)\equiv \frac{a}{\pi} \int^{\infty}_{-\infty}{\frac{\mathrm{e}^{-Y^2}}{(\varv-Y)^{2} + a^{2}} \mathrm{d}Y}.
 \label{Voigt}
\end{equation}
The line opacity is a function of the Voigt profile, and can be calculated by
\begin{equation}
 \chi_\mathrm{l} = \frac{\pi^{1/2} q_\mathrm{e}^{2}}{m_\mathrm{e}c} f_{ij}n_{j} \left( 1 - \frac{g_{j}n_{i}}{g_{i}n_{j}} \right) H(a,\varv),
 \label{line_opacity}
\end{equation}
where $f_{ij}$, $n_i$, $n_j$, $g_i$, and $g_j$ are the oscillator strength between level $i$ and $j$, the populations of the $i$th and 
$j$th levels and the degeneracy of the $i$th and $j$th levels, respectively. The electron mass and charge are represented by 
$m_\mathrm{e}$ and $q_\mathrm{e}$. The damping
constant $\Gamma$ contains the half-width terms for radiative, van der Waals, and Stark broadenings, and has been parameterized by
\citet{Vernazza_etal_1973} as
\begin{eqnarray}
 \Gamma & = & C_\mathrm{rad} + C_\mathrm{vdW} \left( \frac{N_\mathrm{HI}}{10^{16}\,\mathrm{cm^{-3}}} \right)\left( \frac{T}{5000\,
\mathrm{K}} \right)^{0.3} \\ \nonumber 
 & & + C_\mathrm{Stark} \left( \frac{N_\mathrm{e}}{10^{12}\,\mathrm{cm^{-3}}} \right)^{2/3},
\end{eqnarray}
where $C_\mathrm{rad}$, $C_\mathrm{vdW}$ and $C_\mathrm{Stark}$ are the radiative, van der Waals, and Stark half-widths, respectively, in
{\AA}, $N_\mathrm{HI}$ is the number density of neutral hydrogen, and $N_\mathrm{e}$ the number density of electrons.  
Radiative and Stark broadening mechanisms are the most important for the H$\alpha$ line in our modeled environment. Stark broadening is 
significant (or dominant) inside the magnetosphere, due to the high electron densities in this region. However, inside the disk wind, 
$N_\mathrm{e}$ becomes very low, and Stark broadening becomes very weak compared to radiative broadening. These effects are small compared
to the thermal broadening inside the studied region.

The total emerging power in the line of sight for each frequency ($P_{\nu}$) is calculated by integrating the intensity $I_{\nu}$ of 
each ray over the projected area $\mathrm{d}P\,\mathrm{d}Q$. The emerging power in the line of sight depends only on the direction. 
Thus, instead of using flux in our plots, which also depends on the distance from the source, we present our profiles as power plots.
\begin{figure*}[t]
  \centering
  \mbox{\hbox{
  \includegraphics[width=.33\hsize]{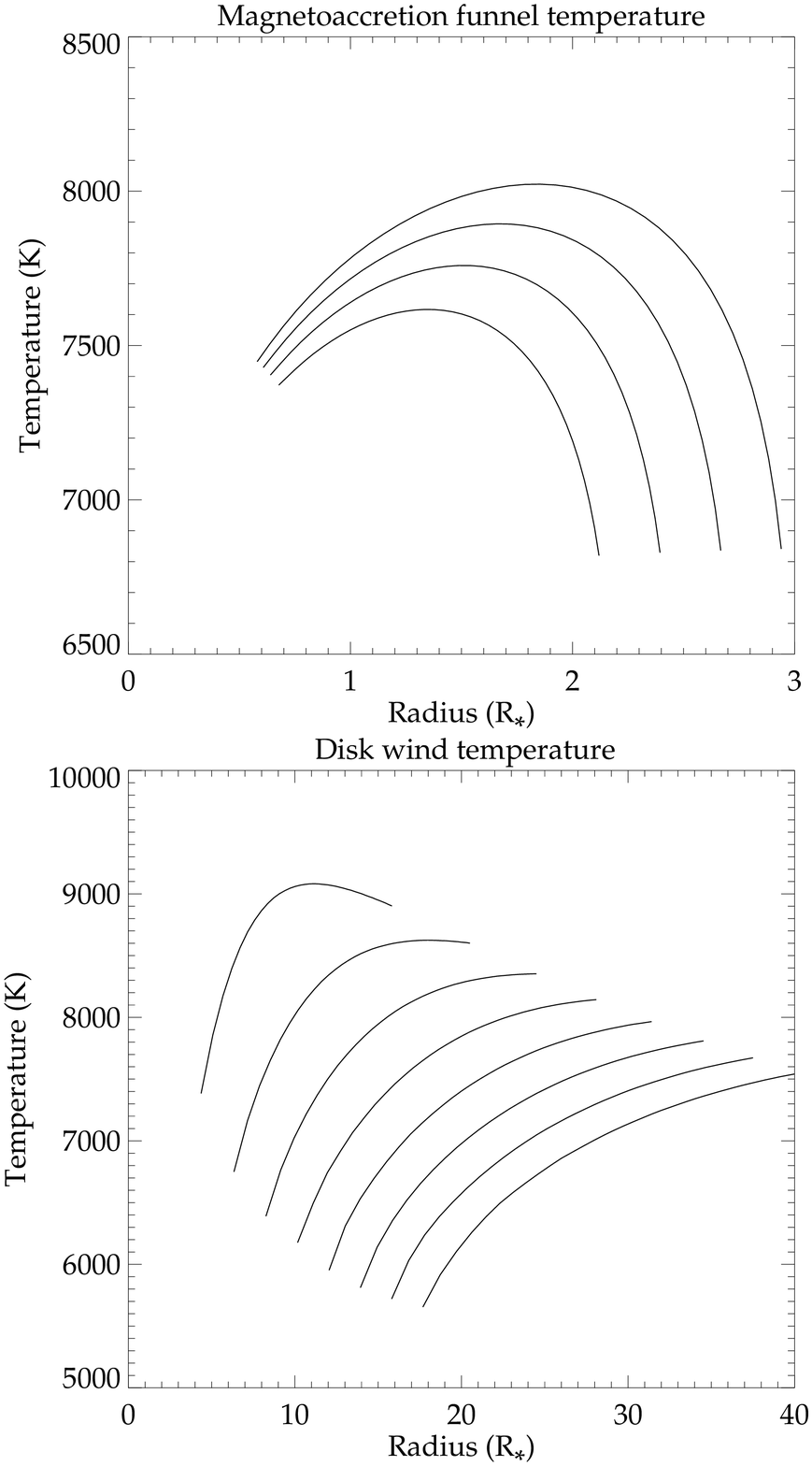}
  \includegraphics[width=.33\hsize]{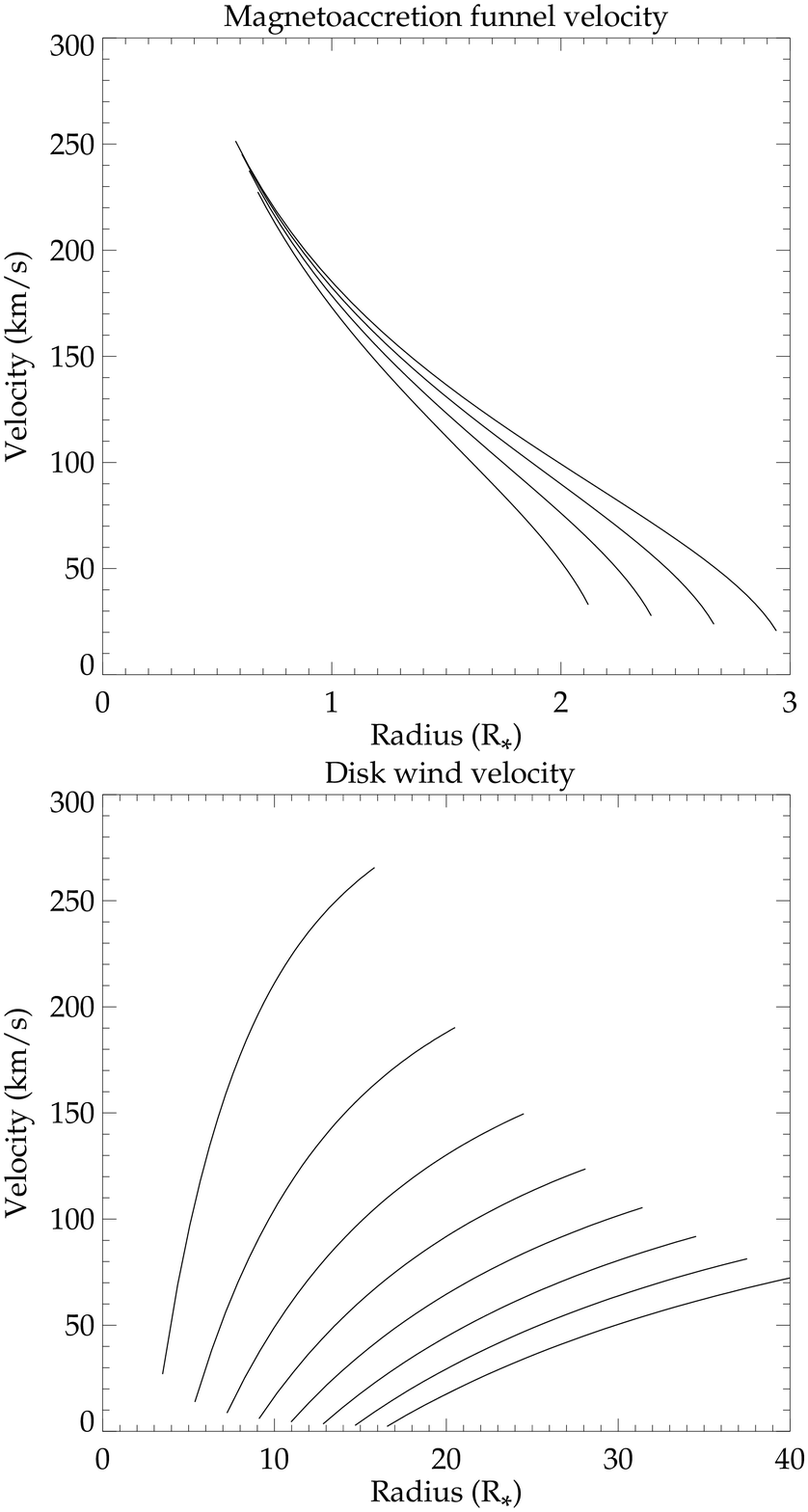}
  \includegraphics[width=.33\hsize]{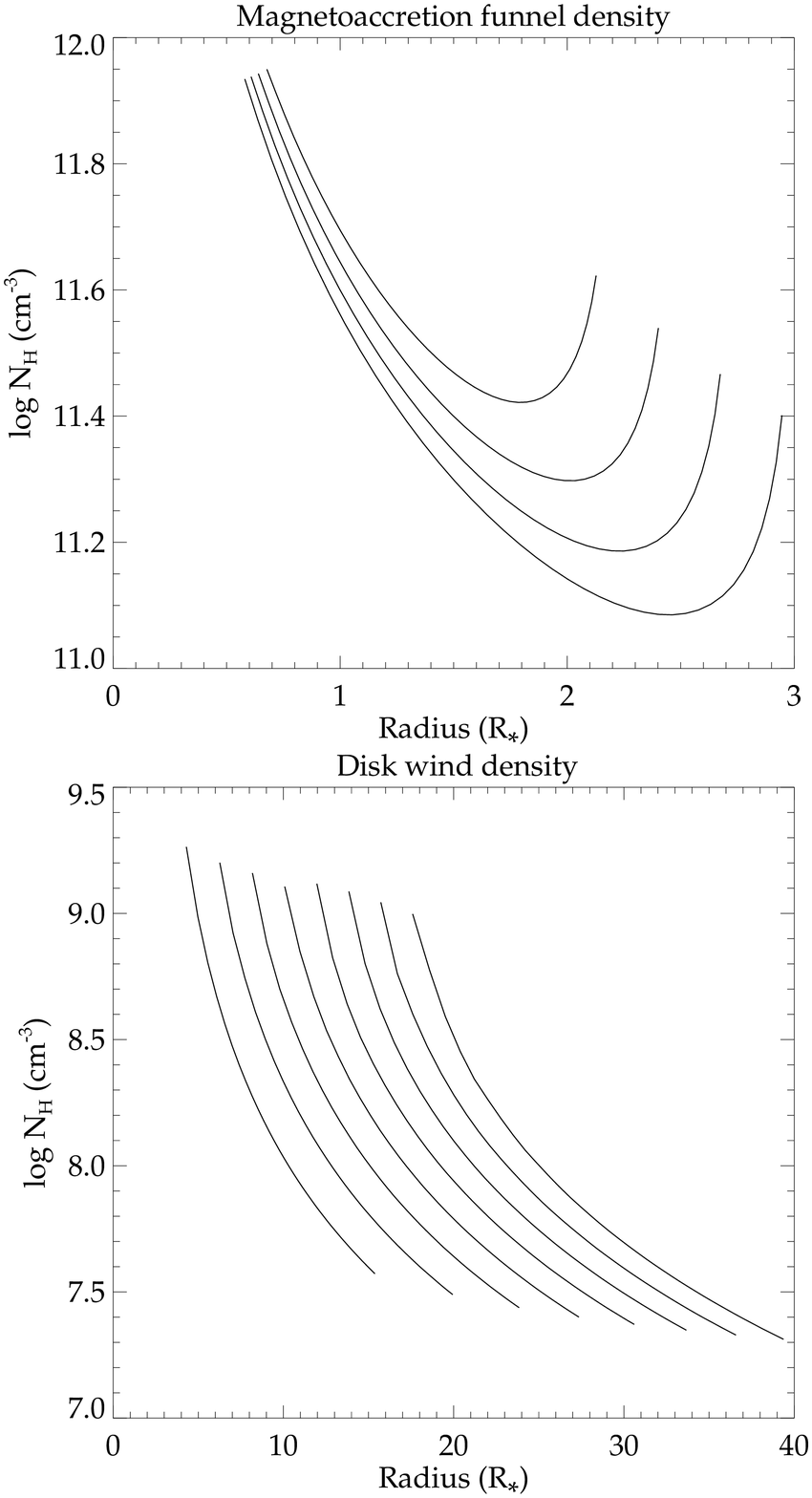}
  }}
 \caption{Temperature (left), poloidal velocity (middle), and density (right) profiles along streamlines inside the magnetosphere (top) 
and along streamlines inside the disk wind region (bottom). In all plots the \emph{x-axis} represents the system cylindrical radius 
in units of stellar radii
$\mathrm{R_*}$. The maximum temperature inside the magnetosphere is \mbox{$T_\mathrm{mag,MAX} \simeq 8000\,\mathrm{K}$}, and 
inside the disk wind it is \mbox{$T_\mathrm{wind,MAX} \simeq 9000\,\mathrm{K}$}. Only some of the streamlines are shown in the plots,
and they are the same ones in the left, middle, and right plots. \label{Temp_magdisk}}
\end{figure*}
\begin{table}[t]
\begin{minipage}[T]{\columnwidth}
 \caption{Model parameters used in the default case. \label{t:tabpar}}
 \centering
 \begin{tabular}{lc|l}
\hline \hline
Stellar Parameters & & Values \\ \hline
Stellar Radius &($R_{*}$) & $2.0 \,\mathrm{R_{\odot}}$ \\
Stellar Mass &($M_{*}$) & $0.5 \,\mathrm{M_{\odot}}$ \\
Photospheric Temperature &($T_\mathrm{ph}$) & $4000 \,\mathrm{K}$ \\[+2pt]
\hline \hline
Magnetospheric Parameters & & Values \\ \hline
Accretion Shock Temperature &($T_\mathrm{sh}$) & $8000 \,\mathrm{K}$ \\
Mass Accretion Rate &($\dot{M}_\mathrm{acc}$) & $10^{-8} \,\mathrm{M_{\odot} \, yr^{-1}}$ \\
Magnetosphere Inner Radius &($r_\mathrm{mi}$) & $2.2 \,\mathrm{R_{*}}$ \\
Magnetosphere Outer Radius &($r_\mathrm{mo}$) & $3.0 \,\mathrm{R_{*}}$ \\[+2pt]
\hline \hline
Disk Wind Parameters & & Values \\ \hline
Inner Radius &($r_\mathrm{di}$) & $3.01 \,\mathrm{R_{*}}$ \\
Outer Radius &($r_\mathrm{do}$) & $30.0 \,\mathrm{R_{*}}$ \\ 
Mass Loss Rate &($\dot{M}_\mathrm{loss}$) & $0.1 \dot{M}_\mathrm{acc}$ \\
Launching angle &($\vartheta_0$) & $33.4 \degr$ \\ 
Dimensionless Angular Momentum & ($\lambda$) & 30.0 \\
Mass Flux to Magnetic Flux Ratio$^a$ & ($\kappa$) & 0.03 \\ [+2pt]
\hline
\multicolumn{3}{l}{$^a$- Dimensionless parameter} \\
%Mass Flux to Magnetic Flux Ratio \footnote{Dimensionless parameter.} & ($\kappa$) & 0.03 \\ [+2pt]
 \end{tabular}
\end{minipage}
\end{table}

\section{Results}

As a starting point, we used the parameters summarized in Table~\ref{t:tabpar} to calculate $\mathrm{H\alpha}$ 
line profiles. We used a default inclination $i$=$55\degr$ in these calculations. For $\dot{M}_\mathrm{loss}$, $r_\mathrm{di}$ and 
$r_\mathrm{do}$ as given by Table~\ref{t:tabpar}, we calculated with Eq.~(\ref{mloss}) a fiducial density $\rho_{0}$=$4.7 \times 
10^{-12}\, \mathrm{g \, cm^{-3}}$. The set of maximum temperatures inside the
magnetosphere and the disk wind region employed in the model ranges from $6000\,\mathrm{K}$ to $10\,000\,\mathrm{K}$, as used by
\citet{MCH2001}. As an example, Fig.~\ref{Temp_magdisk} shows the temperature profiles for the case when the maximum 
temperature inside the magnetosphere, $T_\mathrm{mag,MAX} \simeq 8000\,\mathrm{K}$, and the maximum temperature inside the disk wind,
$T_\mathrm{wind,MAX} \simeq 9000 \,\mathrm{K}$. Figure~\ref{Temp_magdisk} shows some of the streamlines to illustrate the temperature
behavior along the flow lines. Inside the magnetosphere, we can see that there is a temperature maximum midway between 
the disk and the star, and the maximum temperature is higher for streamlines starting farther away from the star. Inside the disk wind
it is the opposite, the maximum temperature along a flux line starts higher near the star, and decays with receding launching 
point from the star. Also for the nearest flux lines the temperature reaches a maximum and then starts falling inside the disk wind, 
and for more extended launching regions the temperature seems to reach a plateau and then remains constant along the line. The
right plots in Fig.~\ref{Temp_magdisk} show the density profiles along the streamlines inside the magnetosphere and disk wind. The 
temperature law used assumes that $T \propto N_{H}^{-2}r^{-3}$, where $N_H$ is the total number density of hydrogen, which 
explains the similarities between the temperature and density profiles. Each streamline is composed of 40 grid points. These lines reach 
a maximum height of $32\,\mathrm{R_*}$ in the disk wind.

Figure~\ref{Temp_magdisk} also shows the poloidal velocity profiles inside the magnetosphere and the disk wind. Inside the magnetosphere,
the velocity reaches its maximum when the flow hits the stellar surface, and it is faster for lines that start farther away from the star.
Inside the disk wind, the outflow speed increases continuously and monotonically, as required by the BP solution, and this acceleration 
is fastest for the innermost streamlines. We can see that the terminal speeds of each streamline inside the magnetosphere have a very low
dispersion and reach values between $\simeq\! 230\,\mathrm{km\,s^{-1}}$ and $\simeq\! 260\,\mathrm{km\,s^{-1}}$. Inside the disk wind, this
velocity dispersion is much higher, with the flow inside the innermost streamline reaching speeds faster than $250\,\mathrm{km\,s^{-1}}$, 
while the flow inside the outermost line shown in Fig.~\ref{Temp_magdisk} -- which is not identical with the outermost line we have used in our default 
case -- is reaching maximum speeds around $\simeq\! 70\,\mathrm{km\,s^{-1}}$. Note that even considering a
thin disk, our disk wind and magnetosphere solutions do not start at $z$=$0$, and the first point in the disk wind streamlines is always at
some height above the accretion disk, which is the reason why the poloidal velocities are not starting from zero.

\begin{figure*}[!t]
 \begin{center}
  \hbox {\includegraphics[width=.5\hsize]{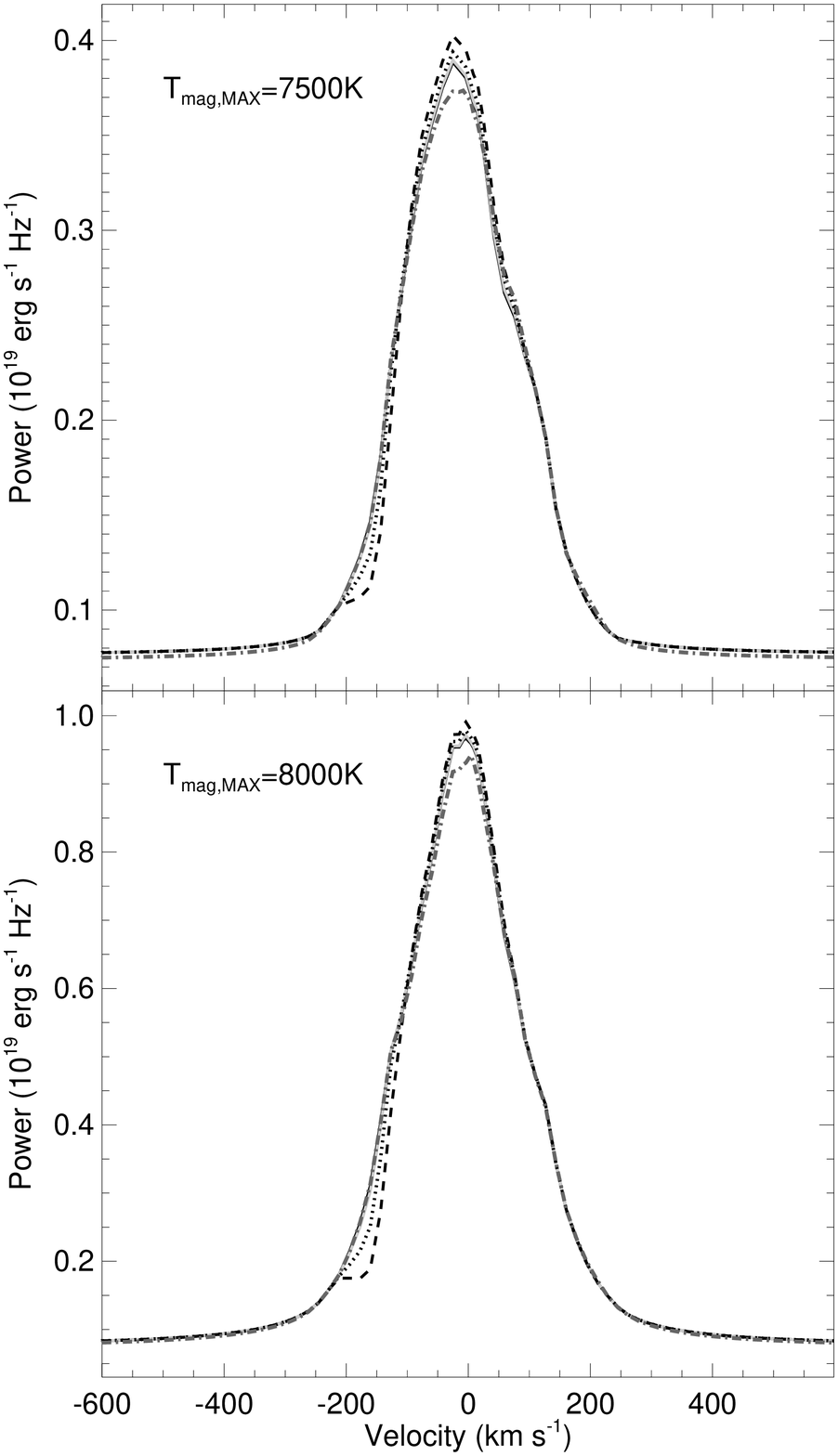}
  \includegraphics[width=.5\hsize]{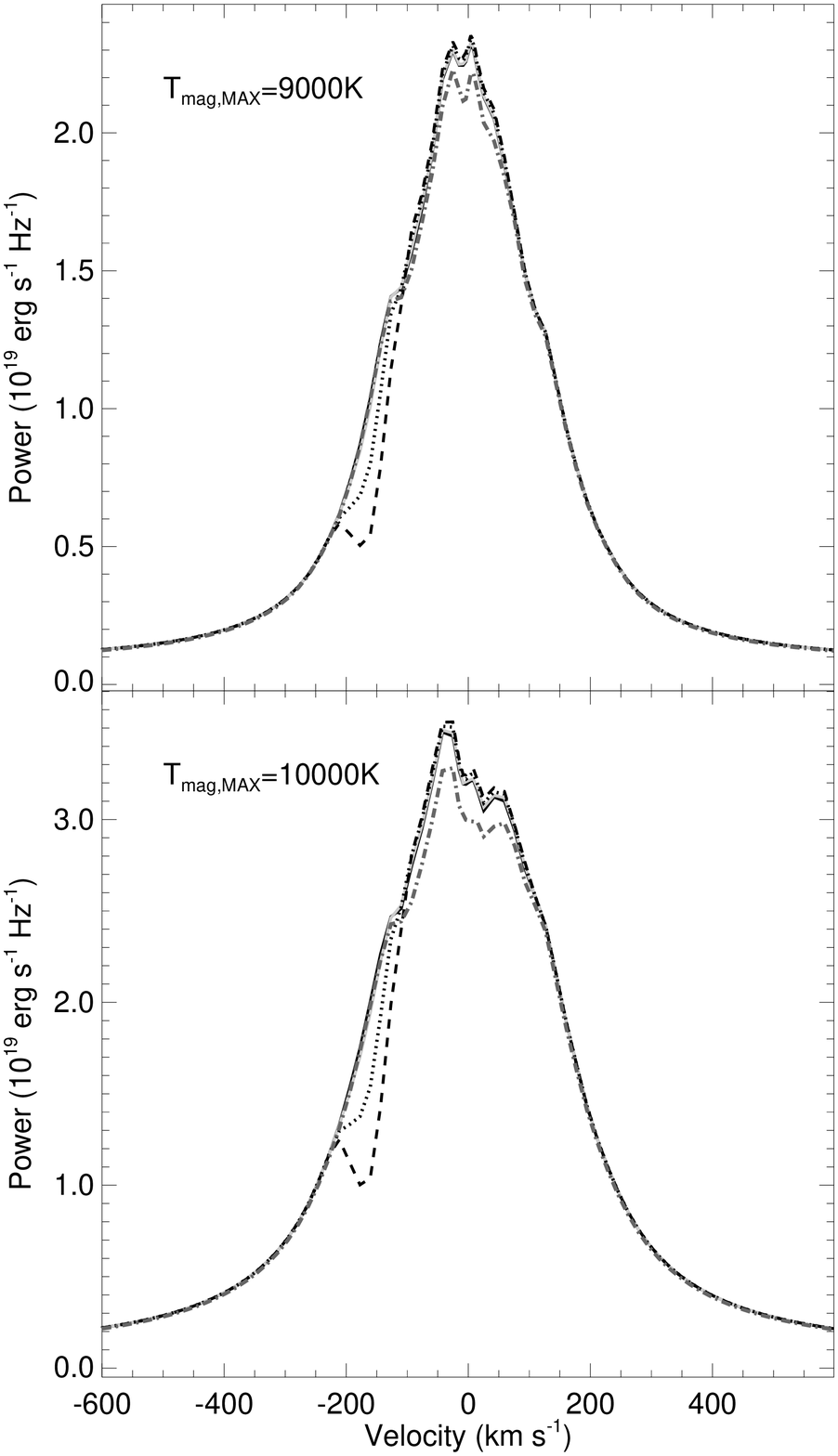}}
 \end{center}
\caption{H$\alpha$ profiles for the default case with values of
$T_\mathrm{mag,MAX}$ ranging from $7500\,$K to $10\,000\,$K and values of
$T_\mathrm{wind,MAX}$ ranging from $7000\,$K to $10\,000\,$K. The maximum
magnetospheric temperature $T_\mathrm{mag,MAX}$ in each plot is the same,
and each different line represents a different maximum disk wind
temperature $T_\mathrm{wind,MAX}$: $7000\,\mathrm{K}$ (black solid line),
$8000\,\mathrm{K}$ (grey solid line), $9000\,\mathrm{K}$
(black dotted line) and $10\,000\,\mathrm{K}$ (black dashed line). The dash-dotted dark grey lines in each plot show the 
profiles calculated with only the magnetospheric component (no disk wind).The black and grey
solid lines completely overlap each other.
% \caption{H$\alpha$ profiles for the default case with values of $T_\mathrm{mag,MAX}$ ranging from $7500\,$K to $10\,000\,$K and values of 
%$T_\mathrm{wind,MAX}$ ranging from $7000\,$K to $10\,000\,$K. The maximum magnetospheric temperature $T_\mathrm{mag,MAX}$ in each plot is 
%the same, and each different line represents a different maximum disk wind temperature $T_\mathrm{wind,MAX}$: $7000\,\mathrm{K}$ 
%(red dot-dot-dot-dashed line), $8000\,\mathrm{K}$ (green dash-dotted line), $9000\,\mathrm{K}$ (blue dashed line) and $10\,000\,\mathrm{K}$ 
%(orange dotted line). The solid black lines in each plot show the profiles calculated with only the magnetospheric component (no disk wind).
\label{Mag_Dwind_default}}
\end{figure*}

The H$\alpha$ line profiles have a clear dependence on the temperature structures of the magnetosphere and the disk wind region. 
Figure~\ref{Mag_Dwind_default} shows plots for different values of $T_\mathrm{mag,MAX}$, and each plot shows profiles for different values 
of $T_\mathrm{wind,MAX}$, which are shown as different lines. The black solid lines are the profiles with only the 
magnetosphere's contribution. One can see that in the default case, most of the H$\alpha$ flux comes from the magnetospheric 
region, which is highly dependent on the magnetospheric temperature. In all plots there is no noticeable difference between 
$T_\mathrm{wind,MAX} \simeq\! 7000\,\mathrm{K}$ (red dot-dot-dot dashed line) and $T_\mathrm{wind,MAX} \simeq\! 8000\,\mathrm{K}$ 
(green dash-dotted line), and the disk wind contribution in these cases is very small, only appearing around the line center 
as an added contribution to the total flux, 
while there are no visible contributions in the line wings. When $T_\mathrm{wind,MAX}$ is around 9000$\,$K, a small blue absorption component 
starts appearing, which becomes stronger as both the magnetospheric and disk wind temperatures increase. The same happens for 
the added disk wind contribution to the flux around the line center. 

\begin{figure}[t]
 \begin{center}
  \includegraphics[width=8.8cm]{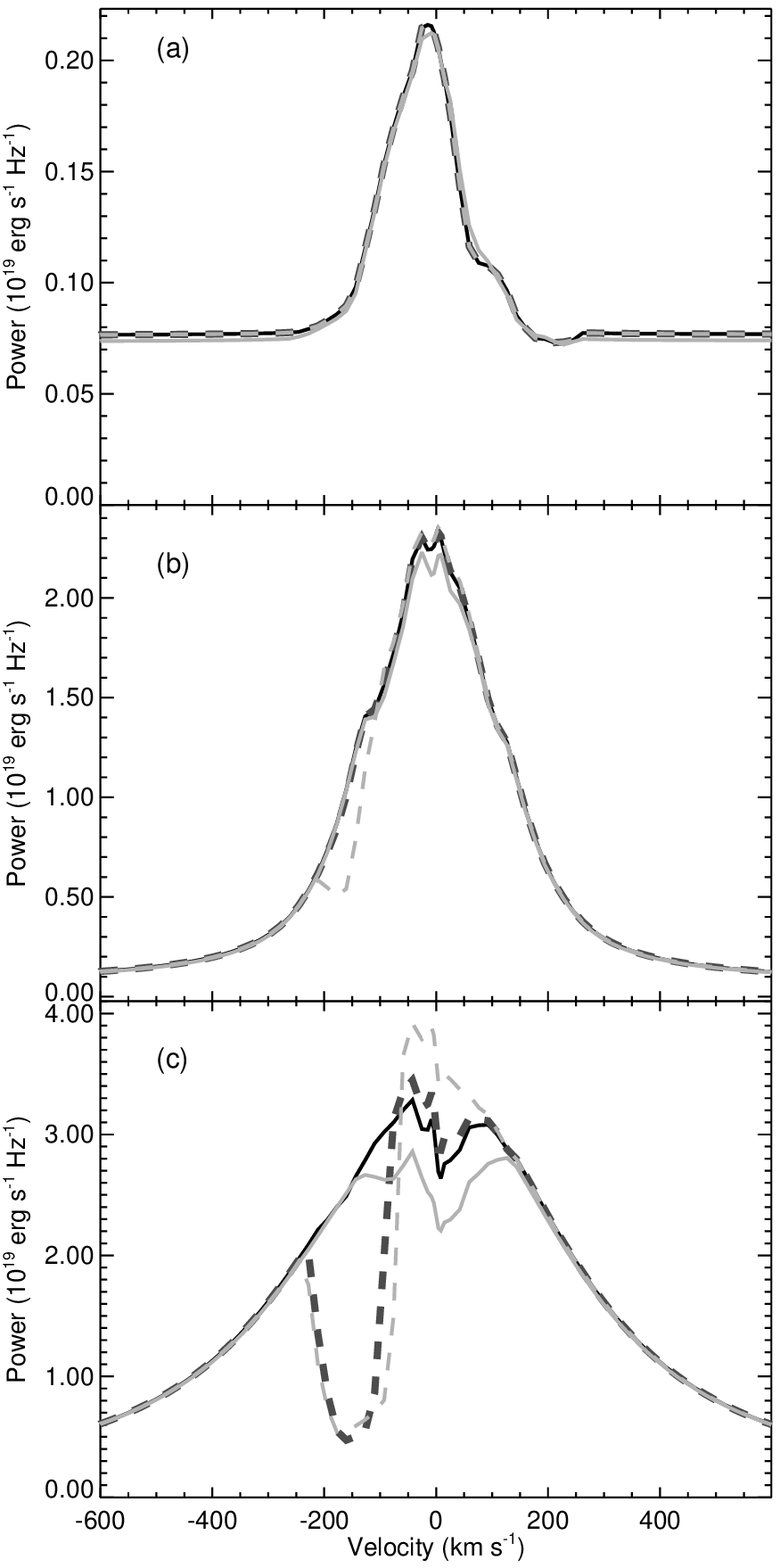}
 \end{center}
\caption{H$\alpha$ profiles for different values of
$\dot{M}_\mathrm{acc}$: (a )$10^{-9}\,\mathrm{M_{\odot}\,yr^{-1}}$,
(b)$10^{-8}\,\mathrm{M_{\odot}\,yr^{-1}}$ and
(c)$10^{-7}\,\mathrm{M_{\odot}\,yr^{-1}}$. The maximum magnetospheric
temperature used is $T_\mathrm{mag,MAX}$$\simeq$9000\,K, and the values
used for $T_\mathrm{wind,MA X}$ are 6000\,K (black thin solid line),
8000\,K (dark grey thick dashed line) and 10\,000\,K (light grey thin
dashed line). The light grey thin solid lines are the profiles where only the
magnetosphere is considered. In all cases, the
$\dot{M}_\mathrm{loss}$$\approx$0.1\,$\dot{M}_\mathrm{acc}$.}
 \label{Mag_Dwind_macc}
\end{figure}

Figure~\ref{Mag_Dwind_macc} shows the variation of the H$\alpha$ line with the mass accretion rate ($\dot{M}_\mathrm{acc}$). In all plots, 
the mass loss rate was kept at $10\%$ of the mass accretion rate, and the maximum temperature inside the magnetospheric accretion funnel was
around $9000\,$K. For $\dot{M}_\mathrm{acc}$=$10^{-9} \,\mathrm{M_{\odot}\,yr^{-1}}$, the plot shows that the line does not change much when the
disk wind temperature varies from $6000\,$K to $10\,000\,$K, because all lines that consider the disk wind contribution are overlapping. There 
is an overall small disk wind contribution to the flux that is stronger around the line center. 
In this case, the disk wind contribution to the line center increases slightly as the temperature becomes hotter than $10\,000\,\mathrm{K}$. 
If the temperature becomes much higher than that, most of the hydrogen will become ionized and no significant absorption in 
H$\alpha$ is to be expected.  
When the mass accretion rate rises to $\dot{M}_\mathrm{acc}$=$10^{-8} \,\mathrm{M_{\odot}\,yr^{-1}}$, the H$\alpha$ flux becomes stronger, 
and the line broader. There is no blue-ward absorption feature until 
the temperature inside the disk wind reaches about $9000\,$K (see Fig.~\ref{Mag_Dwind_default}), after that the blue-shifted absorption 
feature becomes deeper as the temperature rises. Finally, for a system where $\dot{M}_\mathrm{acc}$=$10^{-7}\, \mathrm{M_{\odot}\,
yr^{-1}}$, the line becomes much broader, and even for maximum temperatures inside the disk wind below $8000\,$K, a blue-shifted 
absorption component can be seen. This absorption becomes deeper with increasing disk wind temperature. In this last case, the disk 
wind contribution to the flux is on the same order of magnitude as the contribution from the magnetosphere, which does not happen 
for lower accretion rates. 

\begin{figure}[t]
 \begin{center}
  \includegraphics[width=8.8cm]{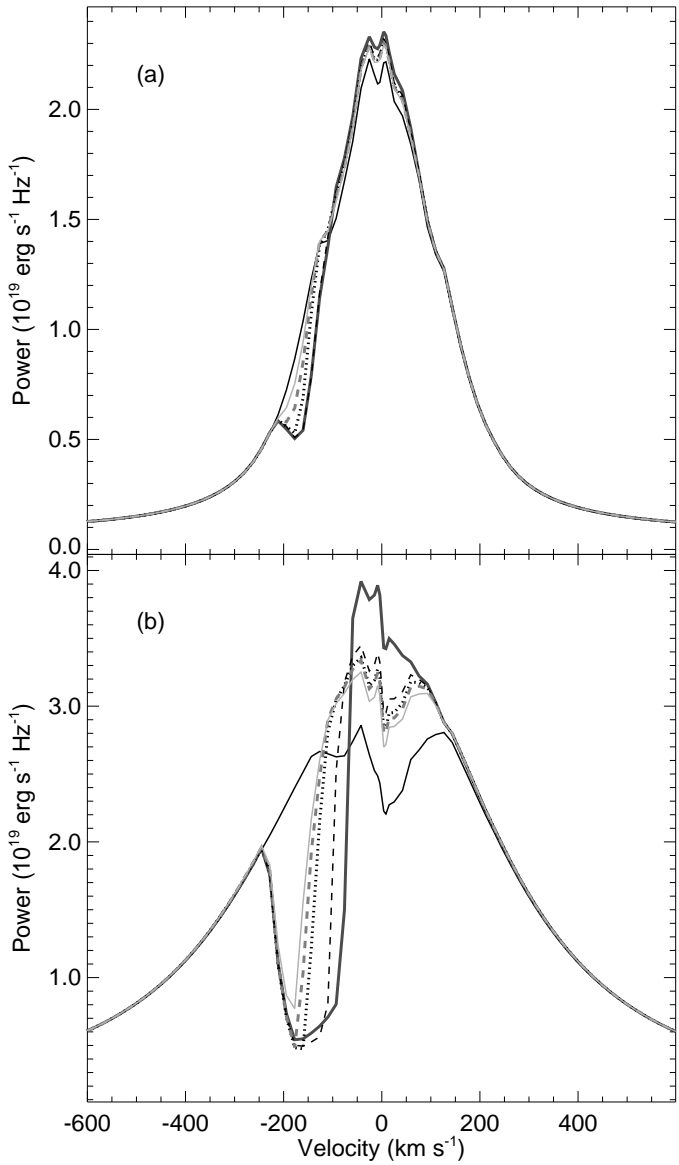}
 \end{center}
\caption{H$\alpha$ profiles for (a)
$\dot{M}_\mathrm{acc}$=$10^{-8}\mathrm{M_{\odot}yr^{-1}}\!$, and (b)
$\dot{M}_\mathrm{acc}$=$10^{-7}\mathrm{M_{\odot}yr^{-1}}\!$, with
$T_\mathrm{mag,MAX}$=9000\,K and $T_\mathrm{wind,MAX}$=10\,000\,K. In all
plots the value of $\rho_0$ is kept constant at a value which makes
$\dot{M}_\mathrm{loss}$$\approx$\,0.1$\,\dot{M}_\mathrm{acc}$, when
$r_\mathrm{do}$=30.0$\,\mathrm{R_{*}}$. The black thin solid lines
are the line profiles with only the magnetosphere; the thick darkgrey
solid lines are the line profiles when magnetosphere and disk wind are
considered and $r_\mathrm{do}$=30.0$\,\mathrm{R_{*}}$. The other
curves show the line profiles when $r_\mathrm{do}$=10.0$\,\mathrm{R_{*}}$
(thin black dashed lines), $r_\mathrm{do}$=5.0$\,\mathrm{R_{*}}$ (thin
black dotted line), \mbox{$r_\mathrm{do} = 4.0 \,\mathrm{R_{*}}$} (dark
grey thin dashed line) and \mbox{$r_\mathrm{do} = 3.5\,\mathrm{R_{*}}$}
(light grey thin solid line).}
 \label{Mag_Dwind_rdo}
\end{figure}

When $\rho_0$ is kept constant and the outer wind radius varies, then $\dot{M}_\mathrm{loss}$ also changes. When that is done, it is also
possible to discover the region inside the disk wind which contributes the most to the line flux. Figure~\ref{Mag_Dwind_rdo} shows
how the line changes with $r_\mathrm{do}$, while $\rho_0$ remains constant. Where \mbox{$\dot{M}_\mathrm{acc}$=$10^{-8}\, 
\mathrm{M_{\odot}\,yr^{-1}}$} (Fig.~\ref{Mag_Dwind_rdo}a), we can see that the line profiles for $r_\mathrm{do}$=$30.0 \,\mathrm{R_{*}}$
and $r_\mathrm{do}$=$10.0 \,\mathrm{R_{*}}$ are almost at exactly the same place, which means that
in this case most of the contribution of the disk wind comes from the wind that is launched at $r\! <\! 10\,\mathrm{R_{*}}$.
And as $r_\mathrm{do}$ becomes even smaller, the blue-shifted absorption feature begins to become weaker and the line peak starts 
to decrease, until the only contributing factor to the line profile is the magnetospheric one. For \mbox{$\dot{M}_\mathrm{acc}$=$10^{-7} \,
\mathrm{M_{\odot}\,yr^{-1}}$} (Fig.~\ref{Mag_Dwind_rdo}b), a similar behavior can be seen, but in this case the plots show the 
loss of a substantial amount of flux when $r_\mathrm{do}$ goes from $r_\mathrm{do}$=$30.0 \,\mathrm{R_{*}}$ to $r_\mathrm{do}$=$10.0 \,\mathrm{R_{*}}$, an
effect mostly seen around the line peak. The same happens at the blue-shifted absorption feature, which begins to weaken when 
$r_\mathrm{do}\! \lesssim\! 15.0 \,\mathrm{R_{*}}$, and weakens faster as $r_\mathrm{do}$ decreases.
In both cases, most of the H$\alpha$ flux contribution from the wind comes from the region where the densities and temperatures are the 
highest, and this 
region extends farther away from the star for higher mass loss rates, as expected. But even for high values of $\dot{M}_\mathrm{loss}$, all
absorption comes from the gas that is launched from the disk in a radius of at most some tens of $\mathrm{R_*}$ from the star. In both cases, 
it seems that a substantial part of the disk wind contribution to the H$\alpha$ flux around the line center comes from a region very near the inner 
edge of the wind, because we still can see a sizeable disk wind contribution around the line center even when $r_\mathrm{do}$=$3.5\,\mathrm{R_*}$.

\begin{figure}[t]
 \begin{center}
  \includegraphics[width=8.8cm]{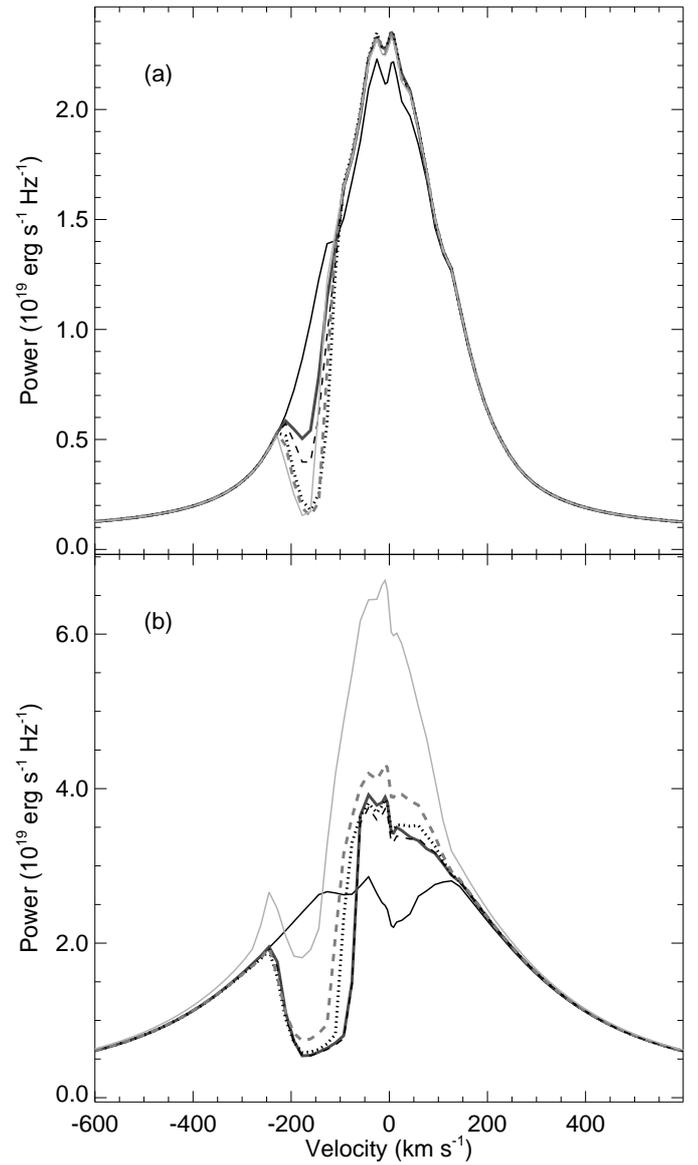}
 \end{center}
\caption{H$\alpha$ profiles as in Fig.\,\ref{Mag_Dwind_rdo}, but now the
density $\rho_0$ is varied to keep the mass loss rate at
$0.1\,\dot{M}_\mathrm{acc}$, while the disk wind outer radius
$r_\mathrm{do}$ changes from $30.0\,\mathrm{R_*}$ to $5.0\,\mathrm{R_*}$,
and each line represents a different $r_\mathrm{do}$ and a different
$\rho_0$. In (a), $r_\mathrm{do}$=30.0$\,\mathrm{R_*}$ and
$\rho_0$=4.71$\times 10^{-12}\mathrm{g\,cm^{-3}}$ (dark grey thick solid
line), $r_\mathrm{do}$=20.0$\,\mathrm{R_*}$ and $\rho_0$=5.72$\times
10^{-12} \mathrm{g\,cm^{-3}}$ (black thin dashed line),
$r_\mathrm{do}$=10.0$\,\mathrm{R_*}$ and $\rho_0$=9.02$\times 10^{-12}
\mathrm{g\,cm^{-3}}$ (black thin dashed dotted line),
$r_\mathrm{do}$=7.5$\,\mathrm{R_*}$ and $\rho_0$=1.19$\times$$10^{-11}
\mathrm{g\,cm^{-3}}$ (dark grey thin dashed line),
$r_\mathrm{do}$=5.0$\,\mathrm{R_*}$ and $\rho_0$=2.13$\times$$10^{-11}
\mathrm{g\,cm^{-3}}$ (light grey thin solid line). In (b), the lines have
the same value of $r_\mathrm{do}$, but their $\rho_0$ are 10 times higher
than in (a). The black thin line represents the line profile with only the
magnetospheric component.}
 \label{Mag_Dwind_rho}
\end{figure}

If instead $\dot{M}_\mathrm{loss}$ is kept constant while both $r_\mathrm{do}$ and $\rho_0$ change, it is possible to infer how the size of the
disk wind affects the H$\alpha$ line as shown in Fig.~\ref{Mag_Dwind_rho}. In the default case (Fig.~\ref{Mag_Dwind_rho}a, 
\mbox{$\dot{M}_\mathrm{acc}$=$10^{-8} \,\mathrm{M_{\odot}\,yr^{-1}}$}), with $T_\mathrm{mag,MAX}\! \simeq\! 9000\,$K,
and $T_\mathrm{wind,MAX}\! \simeq\! 10\,000\,$K, and changing $r_\mathrm{do}$ from $30\,\mathrm{R_{*}}$ to $5\,\mathrm{R_{*}}$ while also changing 
the densities to keep the mass loss rate constant, there are almost no variations around the line peak, as all the profiles with the disk wind
contribution overlap in this region. Some differences can be seen at the blue-shifted absorption component, which 
becomes deeper as the disk wind size decreases and the density increases. For a higher mass loss rate 
(Fig.~\ref{Mag_Dwind_rho}b, \mbox{$\dot{M}_\mathrm{acc}$=$10^{-7} \,\mathrm{M_{\odot}\,yr^{-1}}$}), the line peak decreases when 
$r_\mathrm{do}$ goes from $30\,\mathrm{R_{*}}$ to $20\,\mathrm{R_{*}}$, while the absorption feature barely changes. But as $r_\mathrm{do}$ 
diminishes even further, the line flux as a whole begins to become much stronger again, while the absorption apparently becomes lower. Thus, 
it seems for moderate values of $\dot{M}_\mathrm{acc}$ that the size of the disk-wind launching region does not have a great influence on 
the H$\alpha$ line flux, with only some small differences in the depth and width of the line profile absorption feature. However, for 
higher $\dot{M}_\mathrm{acc}$, the profiles start to vary rapidly when $r_\mathrm{do}\! <\! 20.0 \,\mathrm{R_*}$, and the line
flux increases faster as $r_\mathrm{do}$ decreases.
%For cases with high $\dot{M}_\mathrm{acc}$, the H$\alpha$ lines starts to become more apart from one another for $r_\mathrm{do} < 20.0 \,
%\mathrm{R_*}$, with the line flux increasing faster as $r_\mathrm{do}$ decreases. 
In spite of these differences, we noticed that in both cases this increase 
in the H$\alpha$ flux only becomes important when $\rho_0\! \gtrsim\! 10^{-10} \,\mathrm{g\, cm^{-3}}$, which happens for $r_\mathrm{do}\! \lesssim\! 
7.5\,\mathrm{R_*}$ when $\dot{M}_\mathrm{acc}$=$10^{-7} \,\mathrm{M_{\odot}\,yr^{-1}}$, and for $r_\mathrm{do}\! \lesssim\! 4\,\mathrm{R_*}$, which is out of 
the plot range in Fig.~\ref{Mag_Dwind_rho}a, when $\dot{M}_\mathrm{acc}$=$10^{-8}\,\mathrm{M_{\odot}\,yr^{-1}}$. This indicates that the densities
inside the disk wind region are very important to define the overall H$\alpha$ line shape, if $\rho_0\! \gtrsim\! 10^{-10} \,\mathrm{g\, 
cm^{-3}}$. Still, if $\rho_0$ is smaller, its effect, if any, is smaller and is mostly present around the absorption feature. 

So far we used $\lambda$=$30$ and $\kappa$=$0.03$ in all above mentioned cases. 
The $\lambda$ parameter is the dimensionless angular momentum, and together with $\epsilon$,
the dimensionless specific energy, is constant along a streamline. Equations~(\ref{energy}), (\ref{angular_mom}), (\ref{epsilon}),
and (\ref{lambda}) show that a variation in $\lambda$ will produce a change in the velocities along the streamline.
This change in $\lambda$ will also produce a different disk wind solution, and consequently it is necessary to change the value of $\rho_0$ 
to keep the mass loss rate the same. The effect of $\lambda$ in the line profile can be seen in Fig.~\ref{Mag_Dwind_lambda}a for the 
default model parameters and a constant mass loss rate. It shows that as the value of $\lambda$ 
is increased, the absorption feature becomes deeper and closer to the line center, which means a slower disk wind, and also
an increase in the line flux around its center. Figure~\ref{Mag_Dwind_lambda}b illustrates this behavior and shows how 
the poloidal velocity profile changes when the value of $\lambda$ is varied. 

The H$\alpha$ flux varies in a similar manner when the value of $\kappa$, the dimensionless mass flux to magnetic flux ratio, is changed,
which is illustrated by Fig.~\ref{Mag_Dwind_lambda}c. As $\kappa$ increases, the mass flow becomes slower 
(see Eqs.~\ref{v_scale} and \ref{Eq_f}), and the absorption feature also comes closer to the line center. The higher the value of $\kappa$, 
the deeper is the blue-shifted absorption and the stronger is the flux at the line center, similar to the case when $\lambda$ varies. There is also
a large variation of the poloidal speed inside the flux when $\kappa$ changes, which is shown by Fig.~\ref{Mag_Dwind_lambda}d. Both 
Figs.~\ref{Mag_Dwind_lambda}b and~\ref{Mag_Dwind_lambda}d  confirm the behavior seen in Figs.~\ref{Mag_Dwind_lambda}a 
and~\ref{Mag_Dwind_lambda}c. An increase in the values of $\lambda$ or $\kappa$ decreases the poloidal velocities and 
the velocity dispersion inside the disk wind, causing the absorption feature to approach the line center. The lower velocity dispersion
causes the line opacity to be raised around the absorption feature, which strengthens it. This increase in $\lambda$ or 
$\kappa$ also requires an increase in the densities inside the outflow, if the mass loss rate is to be kept constant, and this effect
also enhances the absorption feature and the line flux around its center. 

The disk-wind launching angle $\vartheta_0$ also produces a variation in the H$\alpha$ line profile. To change $\vartheta_0$, it is
necessary to change the coefficients $a$ and $b$ of Eq.~(\ref{flow_eq}). Here, $b$ is kept constant, and only the coefficient $a$ is varied. 
A change in the coefficients $a$ and $b$ produces another disk wind solution, and then it is necessary to recalculate $\rho_0$  
to hold the mass loss rate constant. In Fig.~\ref{Mag_Dwind_launch}, $\vartheta_0$ varies from $14.6\degr$ to $55.6\degr$, 
which is almost at the launching condition limit, and the value of $a$ used in each case is indicated in the caption.  
Figure~\ref{Mag_Dwind_launch} shows that the absorption feature moves closer to the line center as the launching angle steepens.
Again, as with the cases when we changed $\kappa$ and $\lambda$, a variation in the launching angle produces a variation in the launching 
speed, and this variation is shown at the bottom panel of Fig.~\ref{Mag_Dwind_launch}. The launching speed is slower for higher launching 
angles, and this increases the densities inside the wind. When $\vartheta_0$=$14.6\degr$, the case with the fastest launching speed, 
there is no noticeable difference between the profile with the disk wind 
contribution and the one with only the magnetosphere contribution. This is also the case with the lowest disk wind densities, which
are so low that there is no visible absorption, and just a small disk wind contribution around the center of the line. The case with
$\vartheta_0$=$33.4\degr$ is the default case used in all the prior plots. When the launching angle becomes steeper, the disk wind becomes
denser, and then we have a larger disk wind contribution to the profile and a deeper blue-shifted absorption feature.
Again, the lower velocity dispersion inside the disk wind also helps to enhance the disk wind contribution to the H$\alpha$  
when the launching angle becomes steeper.

One of the most important factors defining the line profiles is the system inclination from the line of sight, $i$.
Figure~\ref{Mag_Dwind_inclination} shows how the H$\alpha$ line changes with $i$, when \mbox{$\dot{M}_\mathrm{acc}$=$10^{-8}\,
\mathrm{M_{\odot}\,yr^{-1}}$}, \mbox{$T_\mathrm{mag,MAX}$=$8000\,$K} and \mbox{$T_\mathrm{wind,MAX}$=$10\,000\,$K}. It is obvious that the profile
intensity is stronger for lower inclinations, and this has a series of reasons. First, because the solid angle containing the studied 
region is larger for lower values of $i$, and this leads to a higher integrated line flux over the solid angle. The opacity inside the accretion
columns is lower if we are looking at them at a lower inclination angle, and then we can see deeper inside, which increases the line flux. 
Finally, the shadowing by the opaque accretion disk and by the star makes the lower accretion columns less visible with a higher $i$,
which also decreases the total flux.  
The H$\alpha$ photon that leaves the magnetosphere or the star crosses an increasingly larger portion of the outflow as $i$ becomes 
steeper, which strengthens the blue-shifted absorption feature. Another effect that happens when the inclination 
changes is that the disk wind velocities projected on the line of sight decrease as the inclination becomes steeper,
and make the blue-shifted absorption feature approach the line center, as is illustrated by Fig.~\ref{Mag_Dwind_inclination}.
The projected velocities increase until $i$ reaches a certain angle, which depends on the disk wind geometry, and
then as the inclination keeps increasing, the projected velocities start to decrease again. The 
disk wind contribution to the red part of the H$\alpha$ profile becomes weaker as $i$ decreases, because there is barely a
contribution to the red part of the profile when $i$=$15\degr$, while this red-shifted contribution seems to be much stronger
when $i$=$60\degr$ and $i$=$75\degr$. If the disk-wind launching angle, $\vartheta_0$, is smaller than $i$, part of the disk wind
flux will be red-shifted. This red-shifted region inside the outflow becomes larger as $i$ increases, and this increases 
the red-ward disk wind contribution to the total line flux. 

The system inclination also affects the continuum level in the line profile. At lower inclinations, the upper accretion ring, 
with its much higher temperature than the stellar photosphere is much more visible than at higher inclinations. Moreover, 
at higher inclinations most of the stellar photosphere is covered by the accretion columns. Then, in a system with low
inclination, the continuum level is expected to be significantly stronger than in systems with a high inclination.  
Analyzing Figs.~\ref{Mag_Dwind_macc} and \ref{Mag_Dwind_inclination} together, we can see that this variation in the
continuum is really significant. Figure~\ref{Mag_Dwind_macc}a shows exactly where the continuum level is when $i$=$55\degr$,
and comparing it with Fig.~\ref{Mag_Dwind_inclination}, it is possible to note an increase between 30\% and 50\% 
in the continuum level when the inclination changes from $75\degr$ to $15\degr$.

Many of the line profiles calculated by our model have some irregular features at their cores. These features become more irregular
as the mass accretion rate and temperature increase. A very high line optical depth inside the magnetosphere due to the increased
density can create some self-absorption in the accretion funnel, resulting in these features. Also, the Sobolev approximation
breakes down for projected velocities near the rest speed. Under SA, the optical depth between two points is inversely proportional 
to the projected velocity gradient between these points. Thus, in subsonic regions like in the base of the disk wind, or in regions
where the velocity gradient is low, SA predicts a opacity that is much higher than the actual one, which creates the irregularities 
seen near the line center that are present in some of the profiles. If instead we do not assume SA, our profiles would be slightly
more intense and smoother around the line center. However, the calculation time would increase considerably.

\section{Discussion}

The analysis of our results indicates a clear dependence of the H$\alpha$ line on the densities and temperatures inside the disk 
wind region. The bulk of the flux comes mostly from the magnetospheric component for standard parameter values, but the disk wind 
component becomes more important as the mass accretion rate rises, and as the disk wind temperature and densities become higher, too. 
For very low mass loss rates (\mbox{$\dot{M}_\mathrm{loss} = 10^{-10} \,\mathrm{M_{\odot}\,yr^{-1}}$}) and disk wind temperatures below 
$10\,000\,$K, there is a slight disk wind component that helps the flux at the center of the line only minimally. No absorption 
component can be seen even for temperatures in the wind as high as $17\,000\,$K, which is expected because in such a hot environment
most of the hydrogen will be ionized. Thus, only the cooler parts of the disk wind where the gas is more rarified would be able
to produce a minimal contribution to the H$\alpha$ flux. The fiducial density in this case is 
\mbox{$\rho_{0} = 4.7 \times 10^{-13} \,\mathrm{g\,cm^{-3}}$}, when the disk-wind launching region covers a ring between $3.0 
\mathrm{R_*}$ ($\approx 0.03\,$AU) and $30.0 \mathrm{R_*}$ ($\approx 0.3\,$AU). Some studies suggest that the outer launching 
radius for the atomic component of
the disk wind is between $0.2 - 3\,$AU \citep{Anderson_etal2003, Pesenti_etal2004, Ferreira_etal2006, Ray_etal2007}, which
is much larger that the values of the disk wind outer radius used here. This means that the densities inside the disk wind should 
be even lower than the values we used, but as the fiducial density falls logarithmically with $r_\mathrm{do}$ (Eq.~\ref{mloss}),
the density would not be much lower. The density for low mass loss rate ($10^{-10} \,\mathrm{M_{\odot}\,yr^{-1}}$) is 
already so low that a further decrease in its value would not 
produce a significant change in the disk wind contribution to the profile. However, the decreased density would create an environment 
where a blue-shifted absorption in the H$\alpha$ line would be much more difficult to be formed. So, for $\dot{M}_\mathrm{acc} < 10^{-9} \,
\mathrm{M_{\odot}\,yr^{-1}}$, the disk wind contribution to the total H$\alpha$ flux is negligible, which is consistent with
observations \citep[e.g.][]{Muzerolle_etal2003, Muzerolle_etal2005}. A blue-shifted absorption in this case is meant to appear if 
the disk wind outer radius is much closer to the star than the observational studies suggest.

For cases with higher values of $\dot{M}_\mathrm{acc}$, and thus higher densities inside the disk wind region, it is possible to see
a transition temperature, below which no blue-shifted absorption feature can be observed (Fig.~\ref{Mag_Dwind_macc}). This
temperature is between $8000\,$K and $9000\,$K when \mbox{$\dot{M}_\mathrm{acc} = 10^{-8} \,\mathrm{M_{\odot}\,yr^{-1}}$},  
between $6000\,$K and $7000\,$K when \mbox{$\dot{M}_\mathrm{acc} = 10^{-7} \, \mathrm{M_{\odot}\,yr^{-1}}$}, and it might be even lower 
than $6000\,$K for cases with very high accretion rates (\mbox{$\dot{M}_\mathrm{acc} > 10^{-7} \,\mathrm{M_{\odot}\,yr^{-1}}$}). 
This transition temperature will directly depend on the mass densities inside the disk wind. Figure~\ref{Mag_Dwind_macc} also shows
that the overall added disk wind contribution to the flux becomes more important as the accretion rate becomes 
stronger, due to a denser disk wind. It is then possible to imagine some conditions that would make the 
disk wind contribution to the H$\alpha$ flux even surpass the magnetospheric contribution: a very high mass accretion rate 
\mbox{($\dot{M}_\mathrm{acc} > 10^{-7} \, \mathrm{M_{\odot}\,yr^{-1}}$)}, or a very large region inside the disk wind with 
temperatures $\sim 9000\,$K, or $r_\mathrm{do} \ll 30\,\mathrm{R_{*}}$, or even a combination of all these conditions.
\begin{figure*}[t]
 \centering
 \mbox{\vbox{
 \hbox{\includegraphics[width=8.8cm]{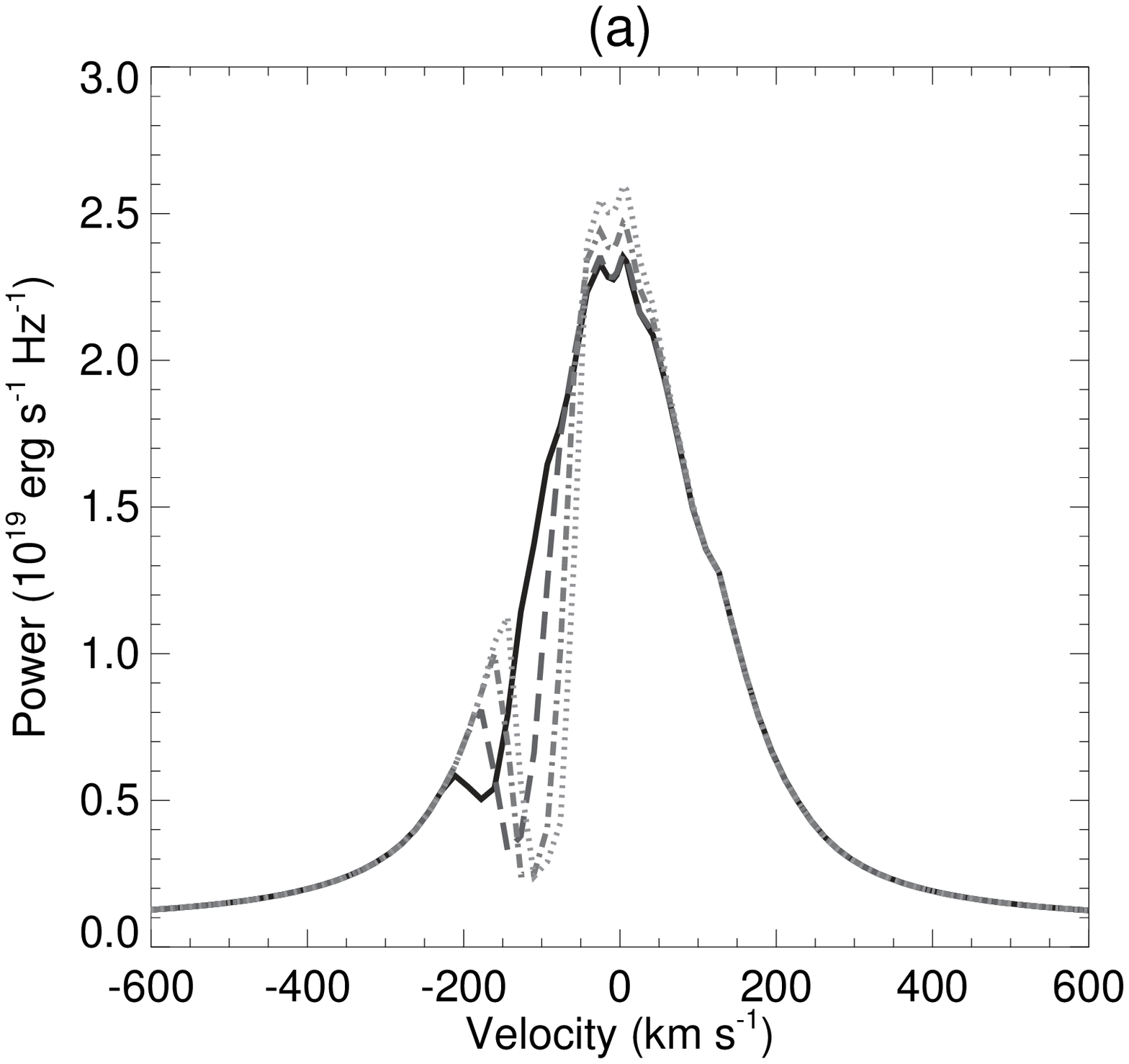}
       \includegraphics[width=8.8cm]{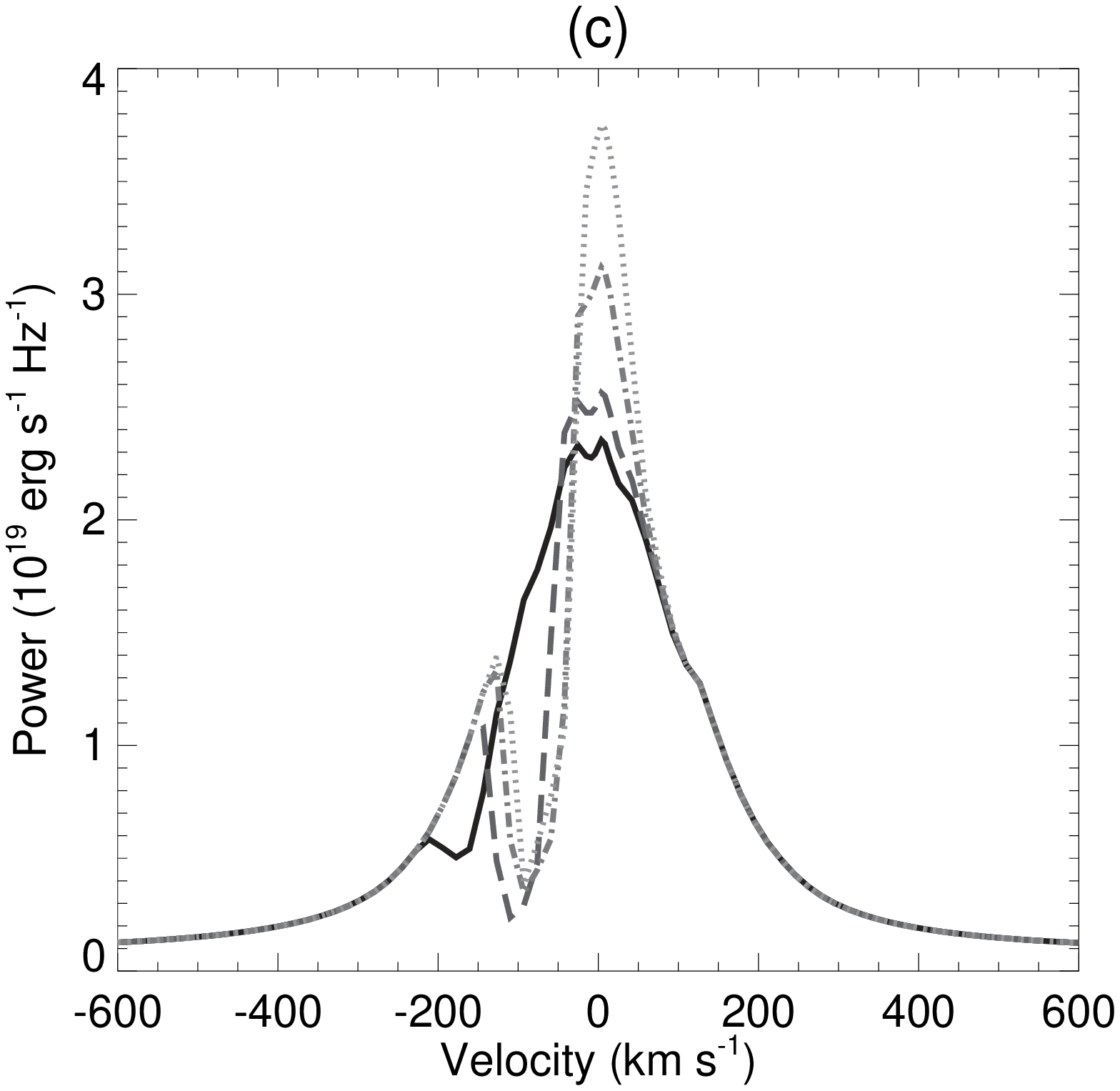}}
 \hbox{\includegraphics[width=8.8cm]{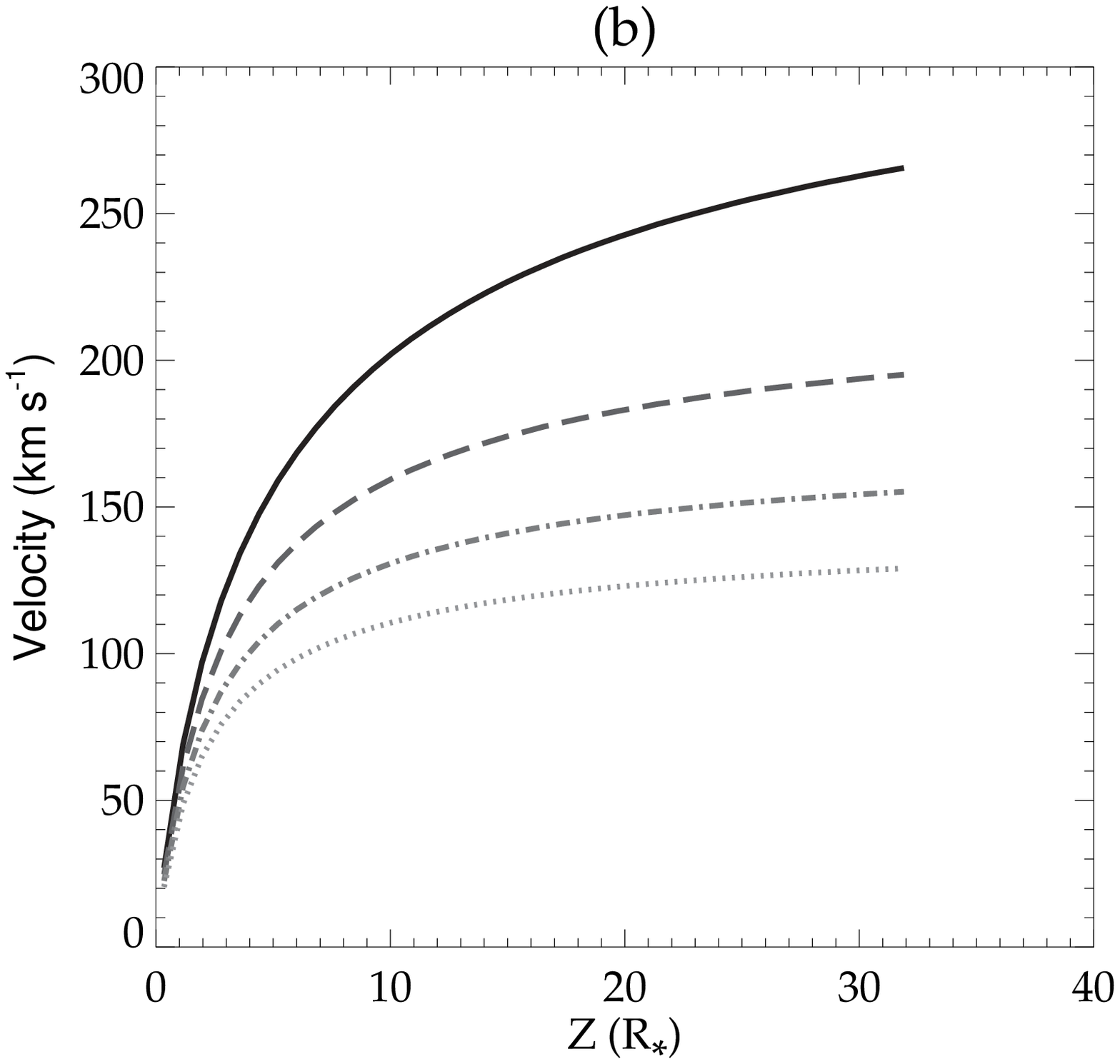}
       \includegraphics[width=8.8cm]{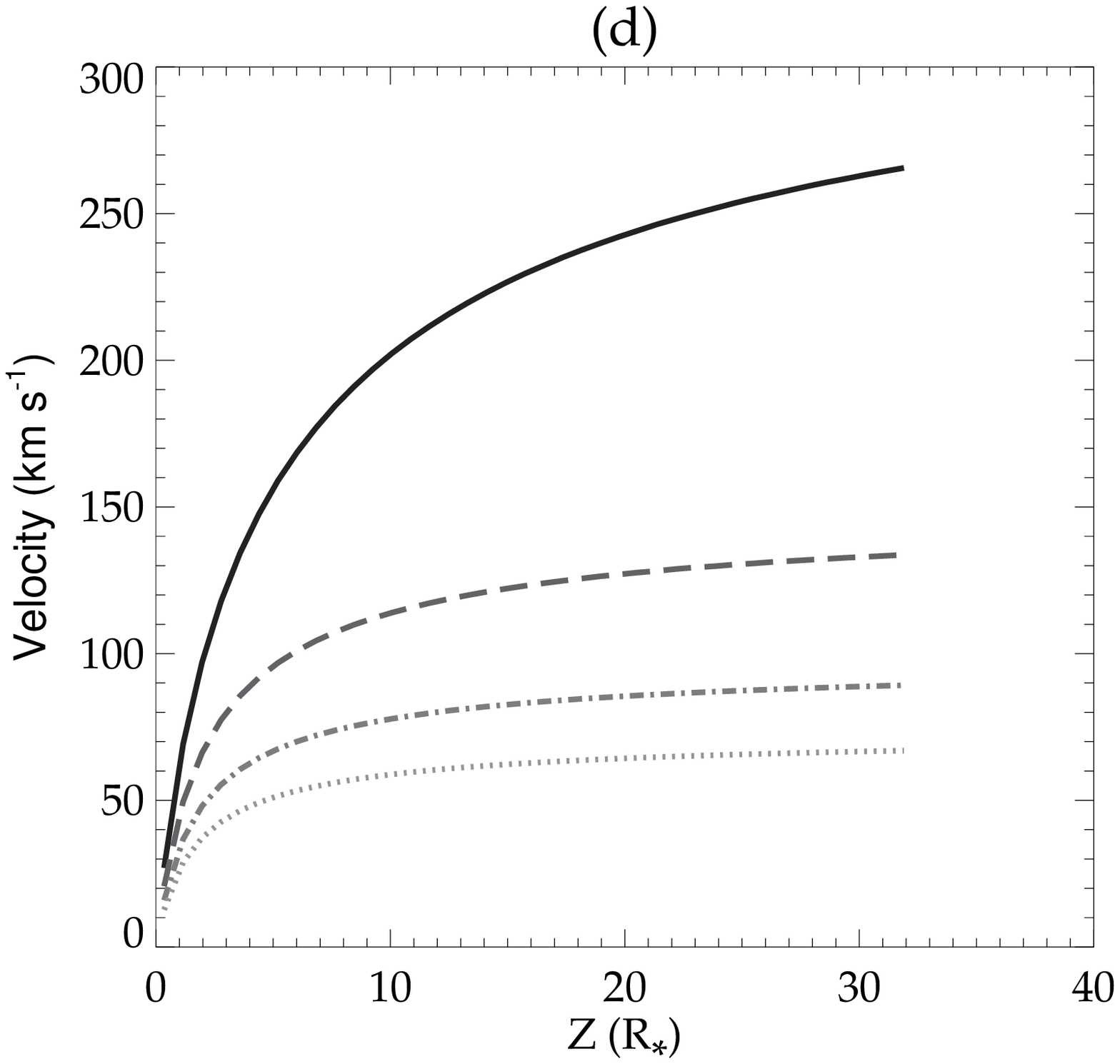}}
 }}
 \caption{H$\alpha$ profiles when \mbox{$\dot{M}_\mathrm{acc} = 10^{-8} \,\mathrm{M_{\odot}\,yr^{-1}}$} with $T_\mathrm{mag,MAX} = 9000\,$K and 
$T_\mathrm{wind,MAX} = 10\,000\,$K. Left panels show (a) the H$\alpha$ profile for different values of $\lambda$, and (b) the corresponding
poloidal velocity profiles for the innermost line of the outflow. The lines in the left panels are: $\lambda=30$ (solid line), $\lambda=40$ 
(dashed line), $\lambda=50$ (dash-dotted line) and $\lambda=60$ (dotted line). The right panels show (c) the H$\alpha$ profile for different
values of $\kappa$, and (d) the corresponding poloidal velocity profiles for the innermost streamline of the outflow. The lines in the right
panel are: $\kappa=0.03$ (solid line), $\kappa=0.06$ (dashed line), $\kappa=0.09$ (dash-dotted line) and $\kappa=0.12$ (dotted line). 
The mass loss rate is kept constant at $0.1 \dot{M}_\mathrm{acc}$. \label{Mag_Dwind_lambda} } 
\end{figure*}

The density inside the outflow alone is a very important factor in the determination of the disk wind contribution to the
H$\alpha$ total flux. The blue-shifted absorption feature that can be seen in many of the profiles is directly affected by
the disk wind densities, which change their depth and width as $\rho_0$ changes. Yet, the other parts of the line profile
seem barely to be affected by the same variation in $\rho_0$, if $\rho_0 \lesssim 10^{-10} \,\mathrm{g\, cm^{-3}}$, and
the mass loss rate and the disk wind solution remain the same. But when $\rho_0 \gtrsim 10^{-10} \,\mathrm{g\, cm^{-3}}$, 
the variation in the line profile begins to adopt another behavior, and the line as a whole becomes dependent on the
disk wind densities, which becomes stronger as $\rho_0$ increases even more. Densities higher than that transition value would
only occur, again, if the mass loss rates are very high, or if $r_\mathrm{do}$ is just a few stellar radii above the stellar surface
in the case of moderate to high mass accretion rates. 

Another interesting result that we obtained shows that even if the disk wind outer radius $r_\mathrm{do}$ is tens of stellar radii 
above the stellar surface, most of the disk wind contribution to the H$\alpha$ flux around line center comes from a region very 
near the disk wind inner edge. This happens because of the higher densities and temperatures in the inner disk wind region if
compared to its outer regions, which makes the inner disk wind contribution much more important to the overall H$\alpha$ flux
than the other regions. This result suggests that the more energetic Balmer lines should be formed in a region narrower than the one 
where the H$\alpha$ line is formed, but a region that also starts at $r_\mathrm{di}$. However, the blue-shifted absorption feature seems 
to be formed in a region that goes from the inner edge
to an intermediate region inside the disk wind, which is $< 30\,\mathrm{R_{*}}$, in the studied cases (see Fig.~\ref{Mag_Dwind_rdo}). 

According to BP, not all values of $\kappa$ and $\lambda$ are capable of producing super-Alfv\'enic flows at infinity, a condition
that enables the collimation of the disk wind and the formation of a jet. The condition necessary for a super-Alfv\'enic flow is
that $\kappa \lambda (2\lambda -3)^{1/2} > 1$, which is true in all plots in this paper. Another condition necessary for a solution is that
it must be a real one and the velocities inside the flux must increase monotonically, which more constrains the set of
values of $\kappa$, $\lambda$, and $\vartheta$, which can produce a physical solution. The BP solution is a ``cold'' wind solution, 
where the thermal effects are neglected and only the magneto-centrifugal acceleration is necessary to produce the outflow from the 
disk. The results presented here show that the disk wind contribution to the H$\alpha$ flux dependents very much on the disk wind 
parameters used to calculate its self-similar solutions. By changing the values of $\kappa$, $\lambda$, and $\vartheta_0$, we change
the self-similar solutions of the problem. Thus, changing the velocities and densities inside the disk wind, leads to a displacement  
of the blue-shifted absorption component and a variation, more visible around the rest velocity, in the overall disk wind contribution to the flux.
For a ``warm'' disk wind solution, with strong heating mechanisms acting upon the accretion
disk surface, and producing an enhanced mass loading on the disk wind, a much lower $\lambda \approx 2 - 20$ would be necessary 
\citep{CasseFerreira2000}. We worked only with the ``cold'' disk wind solution, but it is possible to use a ``warm'' solution, if we
find the correct set of values for $\lambda$ and $\kappa$. 
\begin{figure}[t]
 \centering
 \mbox{\vbox{
 \includegraphics[width=8.8cm]{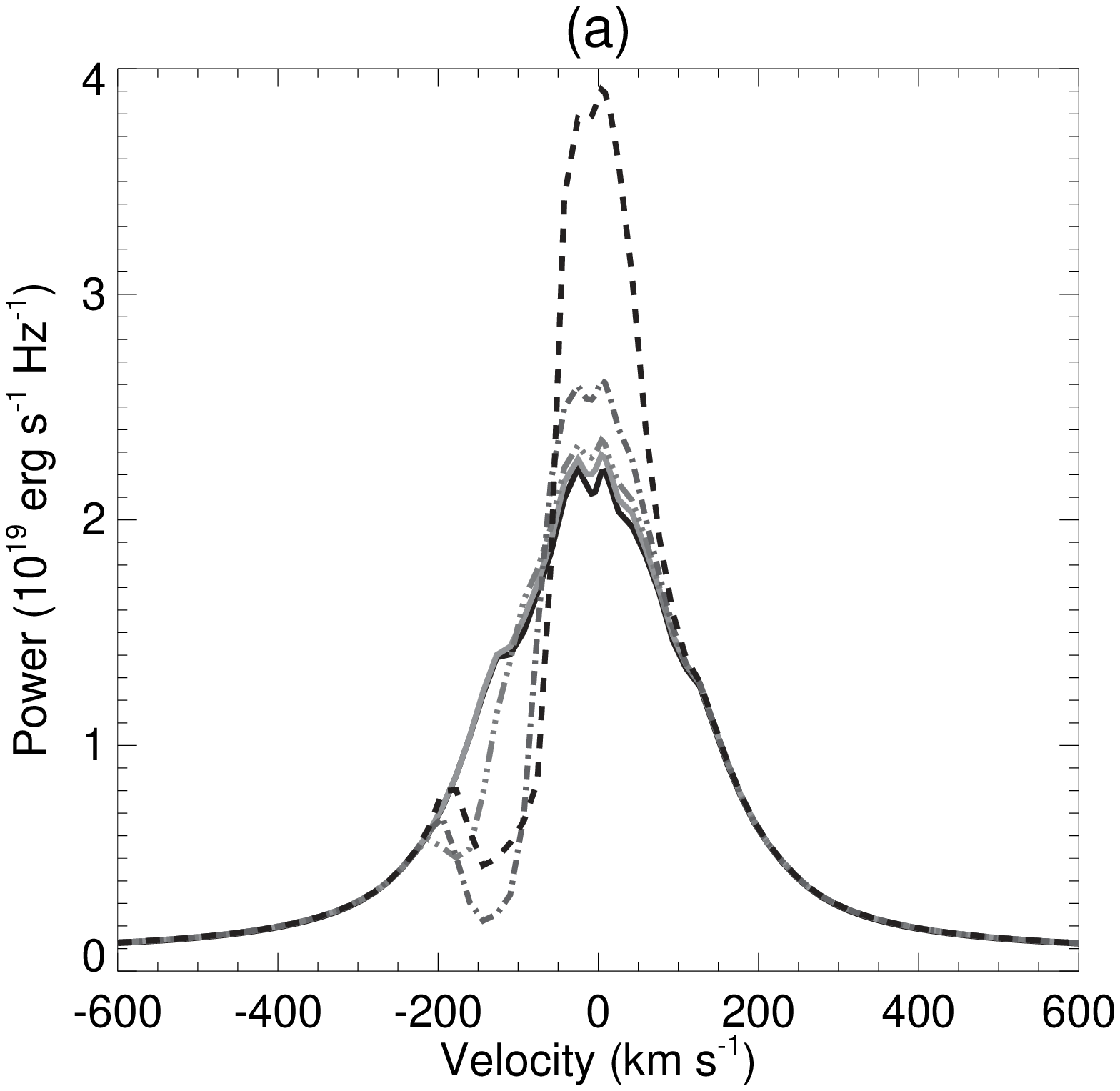}
 \includegraphics[width=8.8cm]{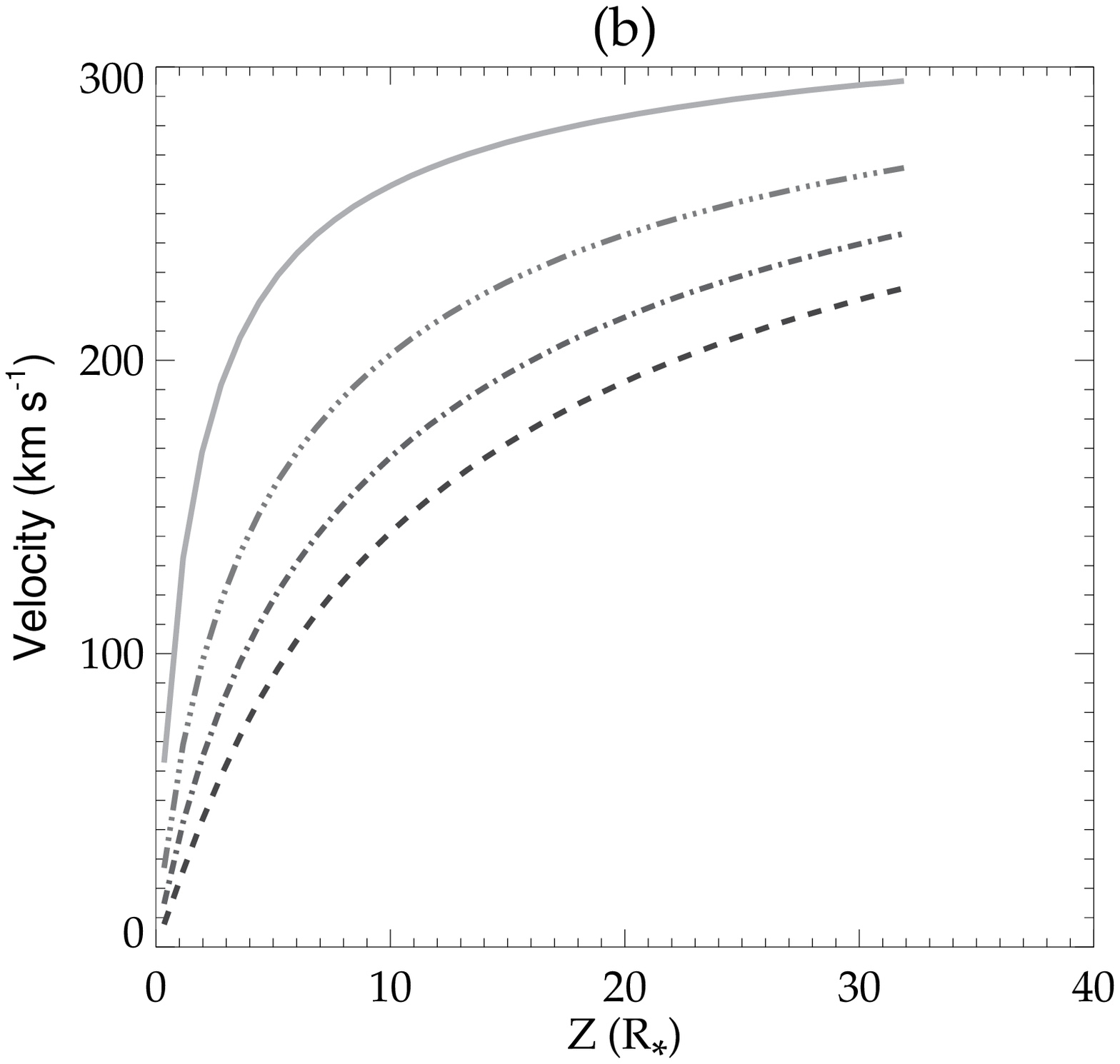}
 }}
 \caption{H$\alpha$ profiles (a) and poloidal velocity profiles for the innermost disk wind streamline (b) for different values 
of the launching angle $\vartheta_0$. The lines are: $\vartheta_0 = 14.6\degr$ and $a = 0.23$ (gray solid line), $\vartheta_0 = 33.4\degr$ 
and $a = 0.43$ (dot-dot-dot-dashed line), $\vartheta_0 = 46.7\degr$ and $a = 0.63$ (dot-dashed line), $\vartheta_0 = 55.6\degr$ and 
$a = 0.83$ (dashed line). The full black line in (a) represents the profile with only the magnetosphere. In all the plots, 
$b=-0.20$, \mbox{$\dot{M}_\mathrm{acc} = 10^{-8}\,\mathrm{M_{\odot}\,yr^{-1}}$}, $T_\mathrm{mag,MAX} = 9000\,$K and $T_\mathrm{wind,MAX} 
= 10\,000\,$K. \label{Mag_Dwind_launch} } 
\end{figure}

Figure~\ref{Mag_Dwind_inclination} shows a dependence of the line profile with inclination $i$ similar to that found by KHS
using their disk-wind-magnetosphere hybrid model. In both models, the H$\alpha$ line profile intensity decreases 
as $i$ increases, and so does the line equivalent width (EW). These results contradict the result from \citet{AppenzellerBertout2005}, 
who have shown that the observed H$\alpha$ from CTTS increase their EW as the inclination angle increases. A solution to that
problem, as shown by KHS, is to add a stellar wind contribution to the models, because that would increase the line EW at lower inclination.
A larger sample than the 12 CTTS set used by \citet{AppenzellerBertout2005} should also be investigated before more definite 
conclusions are drawn. However, there is a great difference between the blue-shifted absorption position in our model
and in the KHS model. In their model, the blue-shifted absorption position is around the line center, while in ours it is
at a much higher velocity. This difference arises from the different disk wind geometry used in both models. \citet{KurosawaHS_2006} used straight 
lines arising from the same source point to model their disk wind streamlines, which results in the innermost streamlines 
having a higher launching angle and thus lower projected velocity along the line of sight in that region when $i$ is large. 
Instead we use parabolic streamlines, and these lines have a higher projected velocity in our default case along the line of sight, 
displacing our absorption feature to higher velocities.

In both models it is common to have the magnetospheric contribution as the major contribution to the H$\alpha$ flux, with a much smaller 
disk wind contribution. In a few of the profiles calculated by KHS, the disk wind contribution is far more 
intense than the magnetospheric contribution. With their disk-wind-magnetosphere hybrid model and using $\dot{M}_\mathrm{acc} = 
10^{-7}\,\mathrm{M_{\odot}\,yr^{-1}}$, $T_\mathrm{wind} = 9000\,\mathrm{K}$ and $\beta=2.0$, KHS have obtained a disk wind contribution 
to the H$\alpha$ flux one order of magnitude higher than the magnetospheric contribution. In their model, $\beta$ is the wind 
acceleration parameter, and the higher its value, the slower is the wind accelerated, which leads to a denser and more slowly  
accelerated outflow in some regions of the disk wind. \citet{KurosawaHS_2006} used an isothermal disk wind. A large acceleration parameter ($\beta=2.0$), 
leading to a large region with a high density, which is coupled with an isothermal disk wind at high temperature ($T_\mathrm{wind} = 9000\,$K), would 
then lead to a very intense disk wind contribution. However, the $\beta=2.0$ value of their model is an extreme condition. For our model to 
produce a disk wind contribution as high as in that extreme case, we would also need very extreme conditions, although due to 
the differences between the models, these conditions should present themselves as a different set of parameters. The modified BP disk wind 
formulation used in our model produces a very different disk wind solution than that used by KHS, the disk wind acceleration 
in our solution is dependent on $\lambda$, $\kappa$ and $\vartheta_0$ parameters (see Figs.~\ref{Mag_Dwind_lambda}b,
~\ref{Mag_Dwind_lambda}d and~\ref{Mag_Dwind_launch}). We also use a very different temperature law (see Fig.~\ref{Temp_magdisk}), 
which is not isothermal, and a different geometric configuration. It is however also possible to obtain a very intense disk wind 
contribution with the correct set of parameters, which corresponds as in KHS to a very dense and hot disk wind solution.

Comparing our results with the classification scheme proposed by \citet{Reipurth_etal1996}, we can see for the H$\alpha$ emission 
lines of TTS and Herbig Ae/Be stars that most of the profiles in this paper can be classified as \emph{Type I} (profiles symmetric 
around their line center) or \emph{Type III-B} (profiles with a blue secondary peak that is less than half the strength of the primary peak).
These two types of H$\alpha$ profiles correspond to almost $60\%$ of the observed profiles in the \citet{Reipurth_etal1996} sample
of 43 CTTS. Figure~\ref{Mag_Dwind_macc}a also shows a profile that can be classified as \emph{Type IV-R}. We have not tried to
reproduce all types of profiles in the classification scheme of \citet{Reipurth_etal1996}, but with the correct set o parameters 
in our model, it should be possible to reproduce them. Even with our different assumptions, our model and KHS model 
also produce many similar results, and both models can reproduce many of the observed types of H$\alpha$ line profiles of CTTS,
but the conditions necessary for both models to reproduce each type of profile should be different.

We are working with a very simple model for the hydrogen atom, with only the three first levels plus continuum. A better
atomic model, which takes into consideration more levels of the hydrogen atom, would give us more realistic line profiles and 
better constraints for the formation of the H$\alpha$ line, and would allow us to make similar studies using other hydrogen 
lines. The assumption that the hydrogen ground level is in LTE due to the very high optical depth to the Lyman lines is 
a very good approximation inside the accretion funnels, and even in the densest parts of the disk wind, 
but it may not be applicable in the more rarified regions of the disk wind. In those regions, the intervening
medium will be more transparent to the Lyman lines, thus making the LTE assumption for the ground level invalid. 
Using a more realistic atom in non-LTE would lower the population of the ground level, and in turn the populations
of levels 2 and 3 would increase. The line source function [Eq.~(\ref{line_source})] depends on the ratio between the populations 
of the upper and lower levels, so, if the population of the level 3 increases more than the population of the level 2, 
the H$\alpha$ emission will increase, otherwise the line opacity will increase. In any case, this will increase the
disk wind effects upon the H$\alpha$ line, thus making the disk wind contribution in cases with low to very low accretion rates
important to the overall profile. The real contribution will depend on each case and is hard to predict.
Still, even with all the assumptions
used in our simple model, we could study how each of the disk wind parameters affects the formation of the H$\alpha$ line on CTTS.
Moreover, the general behavior we found might still hold true for a better atomic model, only giving us different transition 
and limiting values. 

It is necessary to better understand the temperature structures inside the magnetosphere, and inside the disk wind, if we wish to produce 
more precise line profiles. A study about the temperature structure inside the disk wind done by \citet{Safier1993} found that the 
temperature inside the disk wind starts very low ($\lesssim 1000\,$K), for lines starting in the inner $1\,$AU of the disk, 
then reaches a maximum temperature on the order of $10^{4}\,$K, and remains isothermal on scales of $\sim 100 - 1000\,$AU. 
That temperature law is very different from the one used in our models, and it would produce very different profiles
than the ones we have calculated here, because the denser parts of the outflow, in \citet{Safier1993}, have a much lower temperature than
in our model. But \citet{Safier1993} only considered atomic gas when calculating the ionization fraction, and did not consider
H$_2$ destruction by stellar FUV photons, coronal X-rays, or endothermic reactions with O and OH. A new thermo-chemical study
about the disk wind in Class 0, Class I, and Class II objects is prepared by \citet{Panoglou_etal2010}, where they address 
this problem, adding all the missing factors not present in \citet{Safier1993} and also including an extended network of 134 
chemical species. Their results produce a temperature profile that is similar to the ones we used here, but which is 
scaled to lower temperatures. Nevertheless, the innermost streamline they show in their plots is at $0.29\,$AU, which is around 
$30\,\mathrm{R_*}$ in our model and their temperature seems to increase as it nears the star. This strengthens the 
conviction that the temperature structure used in our model is still a good approximation. 

The parameter that defines the temperature structure inside the magnetosphere and inside the disk wind is the $\Lambda$ parameter,
defined by \citet{HartmannAE1982}, and it represents the radiative loss function. It is then possible to use $\Lambda$ to infer the 
radiative energy flux that leaves the disk wind in our models, and even in the most extreme scenarios, the radiative losses in 
the disk wind are much lower than the energy generated by the accretion process. Our model is consequently not violating any energetic
constraints.  

In a following paper we plan to use our model to reproduce the observed H$\alpha$ lines of a set of 
classical T Tauri stars exhibiting different characteristics. After that, we plan to improve the atomic model to obtain more
realistic line profiles of a large variety of hydrogen lines.

\section{Conclusions}

We developed a model with an axisymmetric dipolar magnetic field in the magnetosphere that was firstly proposed by \citet{HHC1994},
and then improved by \citet{MCH1998, MCH2001}, adding a self-similar disk wind component as proposed by \citet{BP1982}, 
and a simple hydrogen atom with only three levels plus continuum. We have investigated which are the most important 
factors that influence the H$\alpha$ line in a CTTS, and which region in the CTTS environment is the most important for the 
formation of this line. We have reached the following conclusions:
\begin{figure}[t]
 \centering
 \includegraphics[width=8.8cm]{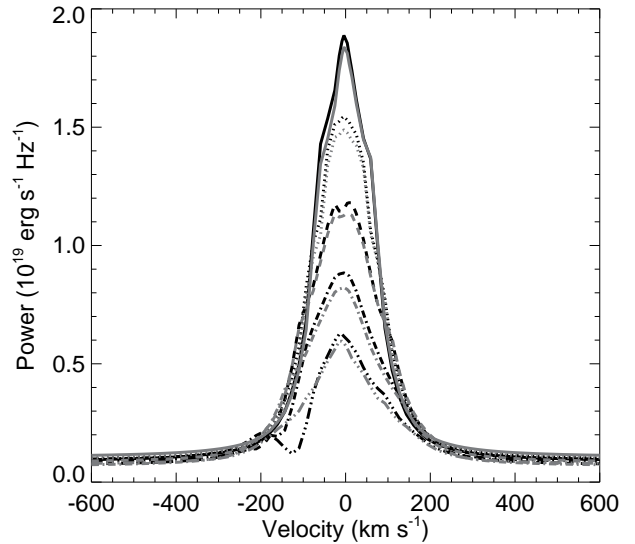}
 \caption{H$\alpha$ profiles with different star-disk system inclinations, $i$, \mbox{$\dot{M}_\mathrm{acc} = 10^{-8}\,\mathrm{M_{\odot}
\,yr^{-1}}$}, $T_\mathrm{mag,MAX} = 8000\,$K and $T_\mathrm{wind,MAX} = 10\,000\,$K. For each $i$, there is a plot showing the total profile
with both magnetosphere and disk wind contributions (black line), and a profile with only the magnetosphere contribution
(gray line). The lines are: $i=15\degr$ (solid), $i=30\degr$ (dotted), $i=45\degr$ (dashed), $i=60\degr$ (dot-dashed), and $i=75\degr$
(dot-dot-dot-dashed). \label{Mag_Dwind_inclination} }
\end{figure}

\begin{itemize}
 \item The bulk of the H$\alpha$ line flux comes from the magnetosphere for CTTS with moderate ($\dot{M}_\mathrm{acc} \approx 10^{-8}\,
\mathrm{M_{\odot}}\,yr^{-1}$) to low mass accretion rates ($\dot{M}_\mathrm{acc} \lesssim 10^{-9}\,\mathrm{M_{\odot}}\,yr^{-1}$). For
CTTS with high mass accretion rates ($\dot{M}_\mathrm{acc} \gtrsim 10^{-7}\,\mathrm{M_{\odot}}\,yr^{-1}$), the disk wind contribution
is around the same order of magnitude as the magnetospheric contribution. In very extreme scenarios, it is possible to make the disk 
wind contribution much more important than the magnetosphere's contribution.

\item There is a transition value of $T_\mathrm{wind,MAX}$ below which no blue-shifted absorption feature can be seen in the H$\alpha$ 
line profile. This transition temperature depends on the mass loss rate, and is lower for higher values of $\dot{M}_\mathrm{loss}$.
This transition value can reach from $T_\mathrm{wind,MAX} \lesssim 6000\,$K for CTTS with very high mass loss rates ($\dot{M}_\mathrm{loss} > 
10^{-8}\,\mathrm{M_{\odot}}\,yr^{-1}$), to $T_\mathrm{wind,MAX} \sim 9000\,$K for stars with moderate mass loss rates ($\dot{M}_\mathrm{loss} \sim
10^{-9}\,\mathrm{M_{\odot}}\,yr^{-1}$). This transition temperature must not be much higher than $10\,000\,$K, otherwise most of the hydrogen
will be ionized. The disk wind in CTTS with \mbox{$\dot{M}_\mathrm{loss} < 10^{-10}\,\mathrm{M_{\odot}}\,yr^{-1}$} should not 
produce any blue-shifted absorption in H$\alpha$. The other disk wind parameters that affect the outflow density can also change the value 
of this transition $T_\mathrm{wind,MAX}$. 

\item Most of the disk wind contribution to the H$\alpha$ flux comes from the inner regions of the disk wind, and the blue-shifted
absorption feature that can be seen in most of the observed profiles is also formed around the same region. For \mbox{$\dot{M}_\mathrm{loss} 
\le 10^{-8}\,\mathrm{M_{\odot}}\,yr^{-1}$}, most of the disk wind H$\alpha$ flux comes from the outflow launched from $r \lesssim 20\,
\mathrm{R_*}$ in the disk. This region is larger for higher mass loss rates. The size of this region depends on the mass loss rate, 
density, and temperatures inside the disk wind.

\item The blue-shifted absorption component dependents very much on the size of the disk wind, but except for this feature, the disk wind
contribution to the H$\alpha$ line barely changes if the $\dot{M}_\mathrm{loss}$ is kept constant while $r_\mathrm{do}$ changes. In
a scenario with a small $r_\mathrm{do}$, the disk wind density might surpass a certain limiting value ($\rho_0 \gtrsim 10^{-10}\,
\mathrm{g\,cm^{-3}}$), making the disk wind contribution to the profile very strong, even the main contribution to the 
profile in more extreme cases. This value for $\rho_0$ should change if we instead use a more realistic hydrogen atom. 

\item The disk wind parameters that define its angular momentum ($\lambda$), mass to magnetic flux ratio ($\kappa$), and launching
angle ($\vartheta_0$) are the main factors that define the position of the blue-shifted absorption feature in the line profile. 
By changing these parameters, it is possible to displace this absorption feature to higher or lower velocities in the blue-ward 
part of the profile.
\end{itemize}

Summarizing, our results indicate that it is necessary to have a substantial outflow ($\dot{M}_\mathrm{loss} > 10^{-9}\,\mathrm{M_{\odot}}\,
yr^{-1}$) from the inner disk ($r \lesssim 0.2 \,\mathrm{AU}$) for the appearance of a visible overall disk wind contribution to
the H$\alpha$ line profile, including the blue-shifted absorption feature. For cases with extreme mass loss rates 
($\dot{M}_\mathrm{loss} > 10^{-8}\,\mathrm{M_{\odot}}\,yr^{-1}$), this region in the inner disk, which is the most important
for the H$\alpha$ line formation, should reach $r \gtrsim 0.2 \,\mathrm{AU}$. The maximum temperature inside the disk wind should also 
be higher than a limiting value, which depends on $\dot{M}_\mathrm{loss}$, for the blue-shifted absorption feature to appear in the 
H$\alpha$ line profile.

We could find better constraints to the model, which would solve the degeneracy encountered in our results, by simultaneously fitting 
other observed hydrogen emission lines of these objects. To do this it is necessary to use a better atomic model. This model can also
be useful to better understand the Herbig Ae/Be stars environment, which, like the CTTS, also produce a diverse amount of emission line 
profiles \citep{Finkenzeller_Mundt1984, Muzeroller_etal2004}. 

\begin{acknowledgements}
The authors thank an anonymous referee for useful suggestions and comments that helped to improve this paper.
GHRAL and SHPA acknowledge financial support from CNPq, CAPES and FAPEMIG. GHRAL also thanks Nuria Calvet, and Lee Hartmann for the 
support during the time he spent as a visiting scholar at the University of Michigan, and Luiz Paulo R. Vaz for
all the valuable discussions, suggestions and help with the plots.
\end{acknowledgements}

\bibliographystyle{aa}
\bibliography{MyBibliography}

\end{document}